\documentstyle[epsfig]{mn}
\input{epsf}

\renewcommand{\d}{{\rm d}}
\newcommand{\beq}{\begin{equation}}
\newcommand{\eeq}{\end{equation}}
\newcommand{\beqa}{\begin{eqnarray}} 
\newcommand{\eeqa}{\end{eqnarray}}
\newcommand{\bea}{\begin{array}} 
\newcommand{\ea}{\end{array}} 
\newcommand{\lag}{\langle}
\newcommand{\rag}{\rangle}
\newcommand{\Om}{\Omega_{\rm m}}
\newcommand{\Ol}{\Omega_{\Lambda}}
\newcommand{\De}{{\cal D}}
\newcommand{\cP}{{\cal P}}
\newcommand{\kappamin}{\kappa_{\rm min}}

\newcommand{\wh}{\hat{w}}
\newcommand{\Map}{M_{\rm ap}}
\newcommand{\bx}{{\bf x}}
\newcommand{\bk}{{\bf k}}
\newcommand{\kpar}{k_{\parallel}}
\newcommand{\kperp}{\bk_{\perp}}
\newcommand{\kperpDt}{k_{\perp}\De\theta_s}
\newcommand{\inta}{\int_{-i\infty}^{+i\infty}}

\newcommand{\Xh}{\hat{X}}
\newcommand{\phiX}{\varphi_{\Xh}}
\newcommand{\Maph}{\hat{\Map}}
\newcommand{\phiMap}{\varphi_{\Maph}}
\newcommand{\wt}{\tilde{w}}
\newcommand{\Mh}{\hat{M}}
\newcommand{\phiM}{\varphi_{\Mh}}
\newcommand{\gammai}{\gamma_i}
\newcommand{\gammais}{\gamma_{i{\rm s}}}
\newcommand{\phigamma}{\varphi_{\hat{\gammais}}}
\newcommand{\Gammai}{\Gamma_i}
\newcommand{\phiG}{\varphi_{\hat{\Gammai}}}
\newcommand{\cM}{{\cal M}}
\newcommand{\cS}{{\cal S}}
\newcommand{\cH}{{\cal H}}
\newcommand{\Pb}{\overline{P}}

\title[Non-Gaussianities and weak lensing surveys]
{On the estimation of gravity-induced non-Gaussianities from weak lensing surveys}
\author[P. Valageas et al.]
{Patrick Valageas$^{1}$, Dipak Munshi$^{2,3}$, Andrew J. Barber$^{4}$\\
$^{1}$Service de Physique Th\'eorique, 
CEA Saclay, 91191 Gif-sur-Yvette, France \\
$^{2}$Institute of Astronomy, Madingley Road,
Cambridge, CB3 OHA, United Kingdom\\
$^{3}$Astrophysics Group, Cavendish Laboratory, Madingley Road, 
Cambridge CB3 OHE, United Kingdom\\
$^{4}$Astronomy Centre, University of Sussex, Falmer, Brighton, BN1 9QJ,
United Kingdom\\
}

\begin{document}
\maketitle

\begin{abstract}
We study various measures of weak lensing distortions in future surveys, taking
into account the noise arising from the finite survey size and the intrinsic
ellipticity of galaxies. We also consider a realistic redshift distribution
of the sources, as expected for the SNAP mission. We focus on the low order
moments and the full distribution function (pdf) of the aperture-mass $\Map$
and of the smoothed shear component $\gammais$. We also propose new unbiased 
estimators for low-order cumulants which have less scatter than the
usual estimators of non-Gaussianity based on the moments themselves. Then,
using an analytical model which has already been seen to provide a good
description of weak gravitational lensing through comparison against numerical
simulations, we study the statistical measures which can be extracted from 
future surveys like the SNAP experiment. We recover the fact that at 
small angular scales ($1'<\theta_s<10'$) the variance can be extracted 
with a few percent level accuracy. Non-Gaussianity can also be measured 
from the skewness of the aperture-mass (at a $10\%$ level) while the shear 
kurtosis is more noisy and cannot be easily measured beyond $6'$. On the 
other hand, we find that the pdf of the estimator associated with the 
aperture-mass can be distinguished
both from the Gaussian and the Edgeworth expansion and could provide useful
constraints, while this appears to be difficult to realize with the shear 
component. Finally, we investigate various survey strategies and the 
possibility to perform a redshift binning of the sample.
\end{abstract}


\section{Introduction}

Weak lensing surveys have already started playing a major role in
cosmology (e.g., Bacon, Refregier \& Ellis, 2000, Hoekstra et al., 2002, 
Van Waerbeke et al., 2001, and Van Waerbeke et al., 2002) not only in 
constraining the background dynamics of the universe but in probing the 
nature of dark matter as well. In order to extract useful information from
observations one needs to compare the data with theoretical predictions
associated with specific cosmological scenarios. To do so, one often uses
numerical simulations which typically employ
ray-tracing techniques as well as line of sight integration of cosmic shear
(e.g., Schneider \& Weiss, 1988, Jaroszynski et al., 1990, 
Wambsganss, Cen \& Ostriker, 1998, Van Waerbeke, Bernardeau \& Mellier, 1999, 
and Jain, Seljak \& White, 2000, Couchman, Barber \& Thomas (1999)). 
On the other hand, several analytical techniques have also been developed 
over the past several
years to predict statistics of weak lensing shear and associated
quantities. On large angular scales perturbative techniques are
generally employed as non-linearities can be treated by a series
expansion (e.g., Villumsen, 1996, Bernardeau et al., 1997, 
Jain \& Seljak, 1997, Kaiser, 1998, Van
Waerbeke, Bernardeau \& Mellier, 1999, and Schneider et al.,
1998). However, on small angular scales, especially relevant to current
observational surveys with small sky coverage, perturbative
calculations are no longer valid and models to represent the
gravitational clustering in the non-linear regime have had to be devised.
One class of such models is based on the hierarchical {\em ansatz} 
(e.g., Fry 1984, Schaeffer 1984, Bernardeau \& Schaeffer 1992, and 
Szapudi \& Szalay 1993, 1997, Munshi, Melott \& Coles 1999) for the evolution 
of high-order correlation functions, joined with Peacock \& Dodds (1996)'s 
prescription (see Peacock \& Smith (2000) for more recent fit) to model the
evolution of the power spectrum or equivalently the two-point correlation 
function. Applications of such hierarchical models
have shown to be quite accurate in predicting various statistics
related to weak lensing shear and convergence at small angular scales
(Valageas 2000a \& b; Munshi \& Jain 2000 \& 2001; Munshi 2000; 
 Bernardeau \& Valageas 2000; Valageas, Barber \& Munshi 2004;
Barber, Munshi \& Valageas 2004; Munshi, Valageas \& Barber 2004).
A second class of theoretical descriptions of the density field is based on 
the halo models (see Cooray \& Seth 2002 for a review) which can also 
reproduce lower order moments (e.g. Takada \& Jain 2002, 
Takada \& Jain 2003a,b).

In order to make contact with observations, in addition to a good description 
of the underlying matter density field, which gives rise to these weak 
lensing effects, one clearly needs to include various sources of noise such 
as the contributions from the intrinsic ellipticity distribution of galaxies, 
shot-noise due to the discreet nature of the source galaxies and finite 
volume effects due to finite survey size. Following Schneider et al. (1998) 
a detailed analysis of these effects was presented in Munshi \& Coles (2003). 
We extend these results in some respects by incorporating a realistic
redshift distribution of sources and generalizing the estimators used
to directly handle the components of shear. We also propose a new family of
estimators for low order moments which have less scatter than usual ones. 
We study in details the influence of the source redshift distribution and
of the intrinsic galaxy ellipticities on the measures, as well as the 
sensitivity to various cosmological or survey parameters.
Extending our analysis to complete probability
distribution functions we investigate to what extent various noise 
contributions make it difficult to distinguish the signatures of
the underlying non-linear dynamics hidden in the tails of the pdf or its shape
near its maximum. In most cases we use for numerical display the parameters 
associated with the SNAP experiment.

This paper is organized as follows: in section~2, we describe weak lensing
observables in general and various filters used to smooth the data, as well as
the estimation of the observational scatter. In section~3, we introduce 
specific survey geometries based on SNAP class experiments and we compute the 
noisy cumulants and the associated probability distribution functions for 
estimators of the shear components as well as the aperture mass. Finally, in 
section~4 we discuss our results.

\section{Statistics of weak-lensing observable}
\label{Statistics}

In a series of papers (Valageas et al. 2004, Barber et al. 2004, 
Munshi et al. 2004) we have described how to obtain the distribution function
of any weak-lensing observables and we have shown that our predictions match
results from numerical simulations for the specific cases of the convergence
$\kappa$, the shear $\gamma$ and the aperture-mass $\Map$. However, in those
studies we did not include the noise associated with the intrinsic ellipticity
of galaxies and we assumed that all sources were located at the same redshift
$z_s$. Here, we show that our formalism can be extended in a straightforward
manner in order to handle these two effects.

\subsection{Redshift distribution of sources}
\label{Redshift distribution of sources}

Let us first recall our notations. Weak-lensing effects can be expressed 
in terms of the convergence along the line-of-sight towards the direction 
${\vec \vartheta}$ on the sky up to the redshift $z_s$ of the source, 
$\kappa({\vec \vartheta},z_s)$, given by (e.g., Bernardeau et al. 1997; 
Kaiser 1998):
\beq
\kappa({\vec \vartheta},z_s) = \frac{3\Om}{2} \int_0^{\chi_s} \d\chi \; 
w(\chi,\chi_s) \; \delta(\chi,\De{\vec \vartheta}) ,
\label{kappazs}
\eeq
with:
\beq
w(\chi,\chi_s) = \frac{H_0^2}{c^2} \; \frac{\De(\chi) \De(\chi_s-\chi)}
{\De(\chi_s)} \; (1+z) ,
\label{w}
\eeq
where $z$ corresponds to the radial distance $\chi$ and $\De$ is the angular 
distance. Here and in the following we use the Born approximation which is
well-suited to weak-lensing studies: the fluctuations of the gravitational
potential are computed along the unperturbed trajectory of the photon 
(Kaiser 1992). Thus the convergence $\kappa({\vec \vartheta},z_s)$ is 
merely the projection of the local density contrast $\delta$ along the 
line-of-sight. Therefore, weak lensing observations 
allow us to measure the projected density field $\kappa$ 
on the sky (note that by looking at sources located at different redshifts 
one may also probe the radial direction). In practice the sources have a broad
redshift distribution which needs to be taken into account. Thus, the quantity
of interest is actually:
\beq
\kappa({\vec \vartheta}) = \int_0^{\infty}\d z_s \; n(z_s) 
\kappa({\vec \vartheta},z_s) , \;\;\; \mbox{with} \; \int\d z_s \; n(z_s)=1 ,
\label{kappanz}
\eeq
where $n(z_s)$ is the mean redshift distribution of the sources 
(e.g. galaxies) normalized to unity. From eq.(\ref{kappazs}), the convergence
$\kappa$ associated with a specific survey also reads:
\beq
\kappa({\vec \vartheta}) = \frac{3\Om}{2} \int_0^{\chi_{\rm max}} \d\chi \; 
\wt(\chi) \; \delta(\chi,\De{\vec \vartheta}) ,
\label{kappa}
\eeq
with:
\beq
\wt(\chi) = \int_z^{z_{\rm max}} \d z_s \; n(z_s) \; w(\chi,\chi_s) ,
\label{wt}
\eeq
where $z_{\rm max}$ is the depth of the survey (i.e. $n(z_s)=0$ for 
$z_s>z_{\rm max}$).
By working with eq.(\ref{kappa}) we neglect the discrete effects due to the 
finite number of galaxies. They can be obtained by taking into account the 
discrete nature of the distribution $n(z_s)$. This gives corrections of 
order $1/N$ to higher-order moments of weak-lensing observables, where $N$ 
is the number of galaxies within the circular field of interest. In practice 
$N$ is much larger than unity (for a circular window of radius 1 arcmin we 
expect $N \ga 100$ for the SNAP mission) therefore in this paper we shall 
work with eq.(\ref{kappa}).

Thus, in order to take into account the redshift distribution of sources we 
simply need to replace $w(\chi,\chi_s)$ in eq.(\ref{kappazs}) by $\wt(\chi)$.
Therefore, all the results of Munshi et al. (2004) remain valid.
Then, usual weak-lensing observables can be written as the angular average 
of $\kappa({\vec \vartheta})$ with some filter $U$:
\beq
X= \int\d{\vec \vartheta}\;U_X({\vec \vartheta})\;\kappa({\vec \vartheta}).
\label{X}
\eeq
For instance, the filters associated with the smoothed convergence $\kappa_s$,
the smoothed shear $\gamma_s$ and the aperture-mass $\Map$ are 
(Munshi et al. 2004):
\beq
U_{\kappa_s}= \frac{\Theta(\vartheta<\theta_s)}{\pi\theta_s^2} , \;\;\; 
U_{\gamma_s} = - \frac{\Theta(\vartheta>\theta_s)}{\pi\vartheta^2}
\; e^{i2\alpha}
\label{Ukappagamma}
\eeq
and
\beq
U_{\Map} = \frac{\Theta(\vartheta<\theta_s)}{\pi\theta_s^2}
\; 9 \left(1-\frac{\vartheta^2}{\theta_s^2}\right) 
\left(\frac{1}{3} - \frac{\vartheta^2}{\theta_s^2}\right) ,
\label{UMap}
\eeq
where $\Theta$ are Heaviside functions with obvious notations and $\alpha$ 
is the polar angle of the vector ${\vec \vartheta}$. The angular radius 
$\theta_s$ gives the angular scale probed by these smoothed observables. Note
that the smoothed shear $\gamma_s$ depends on the matter located outside of
the cone of radius $\theta_s$. However, in practice one directly measures
the shear $\gamma({\vec \vartheta})$ on the direction ${\vec \vartheta}$
(from the ellipticity of a galaxy) and $\gamma_s$ is simply the mean shear 
within the radius $\theta_s$. For $\Map$ we shall use in this paper the filter
(\ref{UMap}), as in Schneider (1996), but one could also use any compensated 
filter with radial symmetry. As in Valageas (2000a), it is convenient to 
define the minimum convergence $\kappamin$ associated with an empty beam 
($\delta=-1$):
\beq
\kappamin = - \frac{3\Om}{2} \int_0^{\chi_{\rm max}} \d\chi \; \wt(\chi) ,
\label{kappamin}
\eeq
and to normalize all observables with respect to $\kappamin$:
\beq
\Xh=\frac{X}{|\kappamin|}=\int_0^{\chi_{\rm max}} \d\chi \; \wh 
\int\d{\vec \vartheta} \; U_X({\vec \vartheta}) \; 
\delta(\chi,\De{\vec \vartheta}) ,
\label{Xh}
\eeq
with:
\beq
\wh=\frac{\wt(\chi)}{\int_0^{\chi_{\rm max}} \d\chi \; \wt(\chi)} .
\label{wh}
\eeq
Then, as described in Munshi et al. (2004), the cumulants of $\Xh$ can be
written as:
\beqa
\lag \Xh^p\rag_c & = & \int_0^{\chi_{\rm max}} \d\chi \; \wh^p 
\int_{-\infty}^{\infty} \prod_{i=2}^{p} \d\chi_i \int \prod_{i=1}^{p} 
\d{\vec \vartheta}_i \; U_X({\vec \vartheta}_i) \nonumber \\ 
& & \times \; \xi_p\left( \bea{l} 0 \\ \De {\vec \vartheta}_1 \ea ,
\bea{l} \chi_2 \\ \De {\vec \vartheta}_2 \ea , \dots , 
\bea{l} \chi_p \\ \De {\vec \vartheta}_p \ea ; z \right) ,
\label{cumXr}
\eeqa
or:
\beqa
\lag \Xh^p \rag_c & = & \int_0^{\chi_{\rm max}} \frac{\d\chi}{2\pi} 
(2\pi\wh)^p \int \prod_{j=1}^{p} \d\bk_{\perp j} \; 
W_X(\bk_{\perp j} \De \theta_s) \nonumber \\ 
& & \times \; \lag \delta(\bk_{\perp 1}) \dots \delta(\bk_{\perp p}) \rag_c .
\label{cumXk}
\eeqa
We note $\lag \dots \rag$ the average over different realizations of the
density field, $\xi_p$ is the real-space $p-$point correlation function of 
the density field $\xi_p(\bx_1,\dots,\bx_p)= \lag \delta(\bx_1) \dots 
\delta(\bx_p)\rag_c$, $\kpar$ is the component of $\bk$ parallel to the 
line-of-sight and $\kperp$ is the two-dimensional vector formed by the 
components of $\bk$ perpendicular to the line-of-sight. In eq.(\ref{cumXk})
we factorized the Dirac term $\delta_D(k_{\parallel 1}+\dots+k_{\parallel p})$
out of the connected correlation 
$\lag \delta(\bk_{\perp 1}) \dots \delta(\bk_{\perp p}) \rag_c$.
We note $W_X(\kperp\De\theta_s)$ the Fourier transform of the window $U_X$:
\beq
W_X(\kperp\De\theta_s) = \int\d{\vec \vartheta} \; U_X({\vec \vartheta}) 
\; e^{i \kperp.\De{\vec \vartheta}} .
\label{WX}
\eeq
In particular, for the smoothed convergence $\kappa_s$, the smoothed shear
$\gamma_s$ and the aperture-mass $\Map$ we have (Munshi et al. 2004):
\beq
W_{\kappa_s}(\kperp\De\theta_s) = \frac{2 J_1(\kperpDt)}{\kperpDt} , 
\label{Wkappa}
\eeq
\beq
W_{\gamma_s}(\kperp\De\theta_s) = \frac{2 J_1(\kperpDt)}{\kperpDt} 
\; e^{i 2\phi} ,
\label{Wgamma}
\eeq
and using eq.(\ref{UMap}):
\beq
W_{\Map}(\kperp\De\theta_s) =  \frac{24 J_4(\kperpDt)}{(\kperpDt)^2} ,
\label{WMap}
\eeq
where $\phi$ is the polar angle of $\kperp$ and $J_{\nu}$ are Bessel functions
of the first kind. The real-space expression (\ref{cumXr}) is well-suited
to models which give an analytic expression for the correlations $\xi_p$,
like the minimal tree-model (Valageas 2000b; Bernardeau \& Valageas 2000; 
Barber et al. 2004) while the Fourier-space expression (\ref{cumXk}) is 
convenient for models which give a simple expression for the correlations
$\lag \delta(\bk_1) .. \delta(\bk_p)\rag_c$, like the stellar model 
(Valageas et al. 2004; Barber et al. 2004). In these two cases one can
resum the cumulants $\lag \Xh^p \rag_c$ which yields the pdf $\cP(\Xh)$ as
(e.g., Munshi et al. 2004):
\beq
\cP(\Xh) = \inta \frac{\d y}{2\pi i \lag\Xh^2\rag_c} \; 
e^{[\Xh y - \phiX(y)] / \lag\Xh^2\rag} ,
\label{PX}
\eeq
where we introduced:
\beq
\phiX(y)= \sum_{p=2}^{\infty} \frac{(-1)^{p-1}}{p!} \; S_p^{\Xh} y^p
\;\;\; \mbox{with} \;\;\; S_p^{\Xh} = \frac{\lag\Xh^p\rag_c}
{\lag\Xh^2\rag^{p-1}} .
\label{phiX}
\eeq
The generating function $\phiX(y)$ is closely related to the 
characteristic function $\varphi(y)$ of the density field. Thus, for the 
stellar model $\phiX(y)$ is merely a suitable average along the line-of-sight
of $\varphi(y)$, while for the minimal tree-model the relationship is slightly
more intricate but explicitly known. For the smoothed convergence, one 
actually has $\varphi_{\hat{\kappa}_{\rm s}}(y)\simeq \varphi(y)$ 
(Valageas 2000a,b; Barber et al. 2004).

\subsection{Intrinsic ellipticity of galaxies, pdf of weak-lensing 
observables}
\label{Intrinsic ellipticity}

\subsubsection{Aperture-mass $\Map$}

Th expressions obtained in the previous section assumed that the observations
were perfect. However, in practice the data exhibits some noise. A specific 
source of noise is merely due to the intrinsic ellipticity of galaxies, which
cannot be avoided. Thus, in order to measure the aperture-mass $\Map$
within a single circular field of angular radius $\theta_s$, in which $N$ 
galaxies are observed at positions ${\vec \vartheta}_j$ with tangential 
ellipticity $\epsilon_{{\rm t},j}$, we can use the estimator $M$ defined by:
\beq
M= \frac{\pi\theta_s^2}{N} \sum_{j=1}^N Q_{\Map,j} \; \epsilon_{{\rm t},j} .
\label{MapQ1}
\eeq
Here we used the fact that the aperture-mass defined from the convergence
$\kappa$ by the compensated filter $U_{\Map}$ given in eq.(\ref{UMap}) can
also be written as a function of the tangential shear $\gamma_{\rm t}$
as (Kaiser et al. 1994; Schneider 1996):
\beq
\Map= \int \d{\vec \vartheta} \; Q_{\Map}({\vec \vartheta}) \; 
\gamma_{\rm t}({\vec \vartheta}) ,
\label{MapQ}
\eeq
with:
\beq
Q_{\Map}({\vec \vartheta}) = \frac{\Theta(\vartheta<\theta_s)}{\pi\theta_s^2}
\; 6  \; \left(\frac{\vartheta}{\theta_s}\right)^2 
\left(1-\frac{\vartheta^2}{\theta_s^2}\right)  .
\label{QMap}
\eeq
In eq.(\ref{MapQ1}) we wrote $Q_{\Map,j}=Q_{\Map}({\vec \vartheta}_j)$. In the
case of weak lensing, $\kappa \ll 1$, the observed complex ellipticity 
$\epsilon$ is related to the shear $\gamma$ by: $\epsilon=\gamma+\epsilon_*$,
where $\epsilon_*$ is the intrinsic ellipticity of the galaxy. Assuming that
the intrinsic ellipticities of different galaxies are uncorrelated random
Gaussian variables, the cumulant of order $p$ of $M$ is:
\beq
\lag M^p\rag_c = \lag \Map^p\rag_c \left( 1+\frac{\delta_{p,2}}{\rho}
\right) \;\;\; \mbox{with} \;\;\; \rho = \frac{2N\lag \Map^2\rag}
{\sigma_*^2 G_{\Map}} ,
\label{Mapc}
\eeq
where $\delta_{p,2}$ is the Kronecker symbol, 
$\sigma_*^2=\lag\epsilon_*^2\rag$
is the dispersion of the intrinsic ellipticity of galaxies and we introduced:
\beq
G_{\Map}= \pi\theta_s^2 \int \d{\vec \vartheta} \; 
Q_{\Map}({\vec \vartheta})^2 .
\label{GMap}
\eeq
For the filter (\ref{QMap}) we have $G_{\Map}=6/5$. In order to obtain 
eq.(\ref{Mapc}) we have averaged i) over the intrinsic ellipticity 
distribution, ii) over the galaxy positions and iii) over the matter
density field, assuming these three averaging procedures are uncorrelated.
The second step can be written for any quantity $X$ as:
\beq
\lag X \rag_{{\rm galaxy} \; {\rm positions}} = \prod_{j=1}^N \int\d z_j \; 
n(z_j) \int \frac{\d{\vec \vartheta}_j}{\pi\theta_s^2} \; X .
\label{galaverage}
\eeq
Thus, since the intrinsic 
ellipticities are Gaussian and we neglected any cross-correlation with the
density field they only contribute to the variance of the estimator $M$.
Note that $M^2$ is a biased estimator of $\lag \Map^2\rag$ because of this
additional term. The quantity $\rho$ measures the relative importance 
of the galaxy intrinsic ellipticities in the signal. They can be neglected if
$\rho \gg 1$. Note that any Gaussian white noise associated with the 
detector can be incorporated into the expression (\ref{Mapc}) by adding
a relevant correction to $\sigma_*^2$. Finally, from eq.(\ref{Mapc}) and
eq.(\ref{phiX}) we obtain for the generating function $\phiM$ of the 
normalized quantity $\Mh=M/|\kappamin|$:
\beq
\phiM(y)= \frac{1+\rho}{\rho} \; \phiMap\left( \frac{\rho}{1+\rho} y \right) 
- \frac{1}{1+\rho} \frac{y^2}{2} .
\label{phiM}
\eeq
Of course, for small $\rho$ we recover the Gaussian (i.e. $\phiM(y)=-y^2/2$)
while for large $\rho$ we recover $\phiMap(y)$.

Thus, each circular field of angular radius $\theta_s$ yields a particular
value for the quantity $M$ defined in eq.(\ref{MapQ1}). If the survey contains
$N_c$ such cells on the sky, we can estimate the pdf $\cP(M)$ through the
estimator:
\beq
P_k = \frac{1}{N_c \Delta} \sum_{n=1}^{N_c} {\bf 1}_k(n) ,
\label{Pk}
\eeq
where ${\bf 1}_k(n)$ is the characteristic function of the interval $I_k$
of width $\Delta$, applied to the value $M(n)$ of $M$ measured in the cell 
$n$:
\beq
{\bf 1}_k(n) = 1 \;\; \mbox{if} \;\; M(n) \in I_k, \;\;\; {\bf 1}_k(n) = 0 
\;\; \mbox{otherwise} ,
\eeq
with:
\beq
I_k= \left[M_k-\frac{\Delta}{2},M_k+\frac{\Delta}{2}\right[ , 
\;\;\;  M_k = k \Delta .
\eeq
For simplicity, we chose all intervals $I_k$ to have the same width $\Delta$,
but this could easily be modified. Note that different intervals do not
overlap and the integer index $k$ runs from $-\infty$ to $+\infty$. 
Then, from the set $\{P_k\}$ we obtain an 
histogram which provides an approximation to $\cP(M)$. Indeed, we have:
\beq
\lag P_k \rag = \Pb_k \;\;\; \mbox{with} \;\;\; \Pb_k = 
\int_{M_k-\Delta/2}^{M_k+\Delta/2} \frac{\d M}{\Delta} \; \cP(M) .
\label{Pbk}
\eeq
Then, for small enough $\Delta$ we have $\Pb_k \simeq \cP(M_k)$. Next, 
assuming that different cells are uncorrelated, the dispersion of the 
estimator $P_k$ is:
\beq
\sigma(P_k)^2= \lag P_k^2 \rag - \lag P_k \rag^2 = \frac{\Pb_k^2}{N_c} 
\left( \frac{1}{\Pb_k\Delta} -1 \right) .
\label{sigPk}
\eeq
Of course, we recover the scaling $\sigma(P_k) \propto 1/\sqrt{N_c}$ where
$N_c$ is the number of cells. On the other hand, since different intervals
$I_k$ do not overlap their covariance is:
\beq
k \neq k' : \;\; \sigma(P_k,P_{k'})^2= \lag P_k P_{k'} \rag - \Pb_k \Pb_{k'} 
= \frac{-1}{N_c} \Pb_k \Pb_{k'} .
\label{cov1}
\eeq

\subsubsection{Smoothed shear component $\gammais$}

In a similar fashion to the aperture-mass, we can measure the shear-component
$\gammai$ (with $i=1,2$) from the estimator $\Gammai$ which we 
define by:
\beq
\Gammai= \frac{\pi\theta_s^2}{N} \sum_j Q_{\gammai,j} \; 
\epsilon_{i,j} .
\label{gammaQ1}
\eeq
Here $\epsilon_{i,j}$ is the component $i$ of the ellipticity of the galaxy
$j$ and for the smoothed shear component $\gammais$ we have 
$Q_{\gammai}({\vec \vartheta}) = 1/(\pi\theta_s^2)$ which is independent 
of ${\vec \vartheta}$. Thus we now get $G_{\gammais}=1$ and we recover
eq.(\ref{phiM}) relating $\phiG(y)$ to $\phigamma(y)$, where we now use 
$G_{\gammais}=1$ for $\rho$ (not to introduce too many notations, we use
the same letter $\rho$ for both the aperture-mass and the shear).

Next, we can estimate the pdf $\cP(\Gammai)$ as in eq.(\ref{Pk}). 
However, we can take advantage of the fact that the pdf $\cP(\Gammai)$ 
is even. Therefore, we can group the intervals $I_{-k}$ and $I_k$ to evaluate 
$\cP(\Gamma_{i,k})$. In other words, we now write:
\beq
P_k = \frac{1}{N_c 2\Delta} \sum_{n=1}^{N_c} {\bf 1}_k(n) , \;\;\; 
\mbox{with} \;\;\; k=0,1,2,..,\infty,
\label{Pkgamma}
\eeq
and:
\beq
I_k= \left[\Gamma_{i,-k}-\frac{\Delta}{2},
\Gamma_{i,-k}+\frac{\Delta}{2}\right[ \cup 
\left[\Gamma_{i,k}-\frac{\Delta}{2},\Gamma_{i,k}+\frac{\Delta}{2}\right[
\eeq
where:
\beq
\Gamma_{i,k} = \left(k+\frac{1}{2}\right) \Delta .
\eeq
This yields:
\beq
\lag P_k \rag = \Pb_k \;\;\; \mbox{and} \;\;\; \sigma(P_k)^2= 
\frac{\Pb_k^2}{N_c} \left( \frac{1}{\Pb_k 2\Delta} -1 \right) ,
\label{sigPkgamma}
\eeq
where $\Pb_k$ is defined as in eq.(\ref{Pbk}). Thus, since $\cP(\Gammai)$
is even we have gained a factor $2$ in the expression (\ref{sigPkgamma}) of
the dispersion $\sigma(P_k)^2$. On the other hand, for $k \neq k'$, the 
covariance is again given by eq.(\ref{cov1}).

Finally, let us note that we kept the term associated with the galaxy 
intrinsic ellipticities, which scales as $1/N$, while we neglected the terms 
associated with the fluctuations of the redshift and angular distribution of 
sources, which also scale as $1/N$. The reason for doing so is that the 
correction due to the galaxy intrinsic ellipticities involves the 
multiplicative factor $\sigma_*^2/\lag \Map^2\rag$ which can be large so that 
$1/\rho$ can be large even though we have $N \gg 1$.

\subsection{Low-order moments}
\label{Low-order moments}

\subsubsection{Aperture-mass $\Map$}

The quantities $M$ and $\gammai$ introduced in the previous section 
provide
biased estimators for the moments of weak-lensing observables. In practice
it is desirable to build unbiased estimators in order to measure low-order 
moments. Thus, as in Schneider et al. (1998) or Munshi \& Coles (2003),
in order to study the aperture-mass we can define the estimators $M_p$ as:
\beq
M_p = \frac{(\pi\theta_s^2)^p}{N(N-1)..(N-p+1)} \sum_{j_1,..,j_p} 
Q_{j_1} \epsilon_{{\rm t},j_1} ... Q_{j_p} \epsilon_{{\rm t},j_p} ,
\label{Mp1}
\eeq
The sum runs over all combinations $\{j_1,..,j_p\}$ with no identical indices.
For simplicity, we wrote $Q$ for $Q_{\Map}$. If we are interested in the 
smoothed shear component $\gammais$ we simply need to use 
$Q_{\gammai,j}$
and $\epsilon_{i,j}$ in eq.(\ref{Mp1}). Then, a straightforward calculation
gives the expectation values of the estimators $M_p$ as well as their 
dispersion $\sigma(M_p)^2$:
\beq
\lag M_p\rag= \lag \Map^p \rag , \;\;\; 
\sigma(M_p)^2 = \lag M_p^2\rag - \lag M_p\rag^2 ,
\label{sigMp}
\eeq
with:
\beq
\sigma(M_1)^2 = \lag \Map^2 \rag \left( 1+\frac{1}{\rho} \right) ,
\label{sigM1}
\eeq
\beq
\sigma(M_2)^2 = \lag \Map^4 \rag_c + 2 \lag \Map^2 \rag^2 
\left( 1+\frac{1}{\rho} \right)^2 ,
\label{sigM2}
\eeq
\beqa
\sigma(M_3)^2 & = & \lag \Map^6 \rag_c + \lag \Map^4 \rag_c \lag \Map^2 \rag
\left( 15+\frac{9}{\rho} \right) + 9 \lag \Map^3 \rag_c^2 \nonumber \\
& & + \lag \Map^2 \rag^3 \left( 15 + \frac{27}{\rho} + \frac{18}{\rho^2}
+ \frac{6}{\rho^3} \right) ,
\label{sigM3}
\eeqa
where $\rho$, which was defined in eq.(\ref{Mapc}), measures the contribution
of the ``cosmic variance'' to the noise, relative to the galaxy intrinsic
ellipticities (and detector white noise). Of course, the dispersion of $M_p$ 
involves the cumulants of $\Map$ up to
order $2p$. In eqs.(\ref{sigM1})-(\ref{sigM3}) we assumed $N \gg 1$ and we
neglected relative corrections of order $1/N$. The estimators $M_p$ correspond
to a single circular field of angular radius $\theta_s$ which contains $N$
galaxies. In practice, the size of the survey is much larger than $\theta_s$
and we can average over $N_c$ cells on the sky. Thus, we define the estimators
$\cM_p$ as:
\beq
\cM_p = \frac{1}{N_c} \sum_{n=1}^{N_c} M_{p,n} ,
\label{cMp}
\eeq
where $M_{p,n}$ is the estimator $M_p$ for the cell $n$. Assuming that these
cells are sufficiently well separated so as to be statistically independent,
we have:
\beq
\lag\cM_p\rag=\lag M_p\rag=\lag \Map^p\rag , \;\;\; \sigma(\cM_p) = 
\frac{\sigma(M_p)}{\sqrt{N_c}} .
\label{sigcMp}
\eeq
Here we assumed for simplicity that all cells have the same number $N$ of 
galaxies. The skewness of the aperture-mass $\Map$ is the coefficient
$S_3^{\Map}$ defined as in eq.(\ref{phiX}): $S_3^{\Map}=\lag \Map^3\rag_c/
\lag\Map^2\rag^2$. Therefore, it can be estimated from the ratio $\cS_3$:
\beq
\cS_3 = \frac{\cM_3}{\cM_2^2} , \;\;\; \lag \cS_3 \rag \simeq S_3^{\Map} ,
\;\;\; \sigma(\cS_3) \simeq \frac{\sigma(\cM_3)}{\lag\Map^2\rag^2} .
\label{cS3}
\eeq
In eq.(\ref{cS3}) we neglected the dispersion of $\cM_2$ in order to obtain
the mean and the dispersion of $\cS_3$. This is not a serious shortcoming
because the error bars increase very fast with the order of the moments so
that the dispersion of $\cS_3$ is dominated by the dispersion of $\cM_3$.
To avoid this approximation one may simply study the cumulants 
$\lag \Map^p\rag_c$ themselves rather than the ratios $S_p^{\Map}$.

\subsubsection{Smoothed shear component $\gammais$}

For the smoothed shear component $\gammais$ we obtain in a similar 
fashion:
\beq
\lag\cM_p\rag=\lag M_p\rag=\lag \gammais^p\rag , \;\;\; \sigma(\cM_p) = 
\frac{\sigma(M_p)}{\sqrt{N_c}} ,
\label{sigcMpgamma}
\eeq
with:
\beq
\sigma(M_2)^2 = \lag \gammais^4 \rag_c + 2 \lag \gammais^2 \rag^2 
\left( 1+\frac{1}{\rho} \right)^2 ,
\label{sigM2gamma}
\eeq
\beqa
\sigma(M_4)^2 & = & \lag \gammais^8 \rag_c + \lag \gammais^6 \rag_c 
\lag \gammais^2 \rag \left( 28+\frac{16}{\rho} \right) 
+ 34 \lag \gammais^4 \rag_c^2 \nonumber \\
& & + \lag \gammais^4 \rag_c \lag \gammais^2 \rag^2 \left( 204
+ \frac{240}{\rho} + \frac{72}{\rho^2} \right) \nonumber \\
& & + \lag \gammais^2 \rag^4 \left( 96 + \frac{240}{\rho} 
+ \frac{216}{\rho^2} + \frac{96}{\rho^3} + \frac{24}{\rho^4} \right) .
\label{sigM4gamma}
\eeqa
Since the shear component $\gammai$ is an even quantity (its sign can be
changed by a simple change of coordinates) all odd moments vanish. The 
kurtosis of the shear $S_4^{\gammais}$ is defined as in eq.(\ref{phiX}):
\beq
S_4^{\gammais} = \frac{\lag\gammais^4\rag_c}
{\lag\gammais^2\rag^3} = \frac{\lag\gammais^4\rag 
- 3 \lag\gammais^2\rag^2}{\lag\gammais^2\rag^3} .
\label{S4}
\eeq
Therefore, it may be estimated from the ratio $\cS_4$:
\beq
\cS_4 = \frac{\cM_4-3\cM_2^2}{\cM_2^3} , \;\;\; \lag \cS_4 \rag \simeq 
S_4^{\gammais} , \;\;\; \sigma(\cS_4) \simeq \frac{\sigma(\cM_4)}
{\lag\gammais^2\rag^3} .
\label{cS4}
\eeq
Here we again neglected the dispersion of $\cM_2$. We must
point out that while the correlation between different cells on the sky
becomes quickly negligible as soon as they do not overlap if we consider the
aperture-mass (Schneider et al. 1998) this is not the case for the shear 
itself. This is due to the fact that $\Map$ involves a compensated filter
which damps the contribution from long wavelengths 
($W_{\Map}(\kperp\De\theta_s) \sim (\kperpDt)^2$ for $\kperpDt \ll 1$ 
in eq.(\ref{WMap})), while the shear is sensitive to low $k$ 
($|W_{\gamma_{\rm s}}(\kperp\De\theta_s)| \rightarrow 1$ for $k \rightarrow 0$
in eq.(\ref{Wgamma})). In this article we shall not investigate this point
and we shall assume that the $N_c$ cells are sufficiently far apart so as to
exhibit a negligible correlation.

\subsection{Improving low-order estimators}
\label{Improving low-order estimators}

The interest of the cumulant $\lag \Map^3\rag_c$ is that it provides a measure
of the deviations from Gaussianity. Moreover, it can be used to break the
degeneracy between the cosmological parameter $\Om$ and the normalization
$\sigma_8$ of the matter power-spectrum (Bernardeau et al. 1997). However, we
can note that the estimator $M_3$ may not be the best way to measure the 
skewness. Indeed, in order to make the most of the departure of the pdf
$\cP(\Map)$ from the Gaussian, we would like to weight the pdf by a factor
which changes sign with the difference $\cP(\Map)-\cP_G(\Map)$, where $\cP_G$
is the Gaussian with the same variance as $\cP$. A detailed study of 
$\cP(\Map)$ was presented in Munshi et al. (2004), see also 
section~\ref{Probability distribution functions} below.
It shows that $\cP(\Map)-\cP_G(\Map)$ does not change sign with $\Map$ (like
$\Map^3$). On the other hand, for small deviations from Gaussianity the pdf
of a quantity $X$ can be expanded as (e.g., Bernardeau \& Kofman 1995):
\beqa
\cP(X) & = & \frac{1}{\sqrt{2\pi\sigma_X^2}} \; e^{-\nu^2/2} \; 
\Bigg \lbrace 1 + \sigma_X \frac{S_3^X}{6} H_3(\nu) \nonumber \\
& & + \sigma_X^2 \left[ \frac{S_4^X}{24} H_4(\nu) + 
\frac{(S_3^X)^2}{72} H_6(\nu) \right] + .. \Bigg \rbrace
\label{Edg}
\eeqa
with:
\beq
\sigma_X = \lag X^2\rag^{1/2} \hspace{0.3cm} \mbox{and} \hspace{0.3cm}
\nu = \frac{X}{\sigma_X} .
\label{nu}
\eeq
Here we introduced the Hermite polynomials $H_n(\nu)$. In particular we have:
\beq
H_3(\nu) = \nu^3 - 3 \nu \hspace{0.3cm} \mbox{and} \hspace{0.3cm}
H_4(\nu) = \nu^4 - 6 \nu^2 + 3 .
\label{Hnu}
\eeq
The Edgeworth expansion (\ref{Edg}) is only useful for moderate deviations 
from the Gaussian, that is when the first correcting term is smaller than 
unity (typically $|\nu| \la 1$ and $|\sigma_XS_3^X|\la 1$). As seen
in Munshi et al. (2004), the Edgeworth expansion (\ref{Edg}) is actually 
not very useful to describe the pdf of the aperture-mass or the shear, since
it only fares well when the pdf is very close to Gaussian. However, as we
have seen in section~\ref{Intrinsic ellipticity}, the noise due to the
intrinsic ellipticity of galaxies makes the pdf of actual observables closer
to Gaussian. Moreover, even if the pdf obtained from eq.(\ref{Edg}) is not
very accurate it gives a reasonable prediction for the sign of the difference
$\cP(\Map)-\cP_G(\Map)$. This leads us to consider for the case of the
aperture-mass the estimators $H_3$ and $\cH_3$ defined by:
\beq
H_3 = M_3 - 3 \cM_2 M_1 , \;\;\; \cH_3=\frac{1}{N_c} \sum_{n=1}^{N_c}H_{3,n},
\label{H3}
\eeq
where $M_p$ and $\cM_p$ are the estimators introduced in 
section~\ref{Low-order moments}. Thus, $H_3$ involves the estimators $M_3$
and $M_1$ associated with the same cell, as well as the mean $\cM_2$ over all
cells. Of course, $H_3$ is built from the Hermite polynomial $H_3(\nu)$
given in eq.(\ref{Hnu}). We see from eq.(\ref{H3}) that it is actually an
estimator for the cumulant $\lag\Map^3\rag_c$ rather than for the moment 
$\lag\Map^3\rag$ like $M_3$ (but in the case of the aperture-mass it happens
that $\lag\Map^3\rag_c=\lag\Map^3\rag$). At lowest order over $1/N_c$ we 
obtain:
\beq
\lag \cH_3\rag=\lag H_3\rag=\lag\Map^3\rag_c , \;\;\; 
\sigma(\cH_3)= \frac{\sigma(H_3)}{\sqrt{N_c}} ,
\label{cH3}
\eeq
with:
\beq
\sigma(H_3)^2= \sigma(M_3)^2 - 6 \lag\Map^4\rag_c \lag\Map^2\rag
- 9 \lag\Map^2\rag^3 \left(1+\frac{1}{\rho}\right) .
\label{sigH3}
\eeq
Therefore, we see that the dispersion of $\cH_3$ is always smaller than
for $\cM_3$. Note that although $H_3$ as written in eq.(\ref{H3}) is biased
by a term of order $1/N_c$, which should not be a problem, this term can be
removed in a straightforward way by replacing $\cM_2$ in eq.(\ref{H3}) by
$(N_c\cM_2-M_{2,n})/(N_c-1)$ (i.e. the mean $\cM_2$ is computed from all 
other $(N_c-1)$ cells). Then, the skewness $S_3^{\Map}$ can be estimated from
\beq
\cS_3^H = \frac{\cH_3}{\cM_2^2} , \;\;\; \lag \cS_3^H \rag \simeq 
S_3^{\Map} , \;\;\; \sigma(\cS_3^H) \simeq \frac{\sigma(\cH_3)}
{\lag\Map^2\rag^2} .
\label{cS3H}
\eeq
In eq.(\ref{cS3H}) we again neglected the dispersion of $\cM_2$.

For the shear, we introduce in a similar fashion the estimators $H_4$
and $\cH_4$ built from the Hermite polynomial of order four:
\beq
H_4 = M_4 - 6 \cM_2 M_2 + 3 \cM_2^2, \;\;\; \cH_4=\frac{1}{N_c} 
\sum_{n=1}^{N_c}H_{4,n} .
\label{H4}
\eeq
This again provides an estimator for the cumulant $\lag\gammais^4\rag_c$
rather than the moment $\lag\gammais^4\rag=\lag\gammais^4\rag_c + 
3 \lag\gammais^2\rag^2$ which was estimated by $M_4$. Then we obtain at 
lowest order over $1/N_c$:
\beq
\lag \cH_4\rag=\lag H_4\rag=\lag\gammais^4\rag_c , \;\;\; 
\sigma(\cH_4)= \frac{\sigma(H_4)}{\sqrt{N_c}} ,
\label{cH4}
\eeq
with:
\beqa
\lefteqn{ \sigma(H_4)^2 = \sigma(M_4)^2 - 12 \lag\gammais^6\rag_c 
\lag\gammais^2\rag } \nonumber \\ 
& & - 12 \lag\gammais^4\rag_c \lag\gammais^2\rag^2
\left(11+\frac{8}{\rho}\right) - 72 \lag\gammais^2\rag^4 
\left(1+\frac{1}{\rho}\right)^2 .
\label{sigH4}
\eeqa
Thus, we note that the dispersion of $\cH_4$ is again always smaller than for 
$\cM_4$. Next, the kurtosis $S_4^{\gammais}$ may be estimated from:
\beq
\cS_4^H = \frac{\cH_4}{\cM_2^2} , \;\;\; \lag \cS_4^H \rag \simeq 
S_4^{\gammais} , \;\;\; \sigma(\cS_4^H) \simeq \frac{\sigma(\cH_4)}
{\lag\gammais^2\rag^2} ,
\label{cS4H}
\eeq
where we again neglected the dispersion of $\cM_2$.

Thus, since the estimators $H_3$ and $H_4$ are no more difficult to use than
$M_3$ and $M_4$ and exhibit a smaller dispersion, they are better suited to
the measure of low-order cumulants. In fact, it can be shown that $H_3$ and
$H_4$ are optimal estimators among their class for a Gaussian distribution.
Thus, if we generalize $H_3$ and $H_4$ to the unbiased estimators $L_3$ 
and $L_4$ defined as:
\beq
L_3= M_3 - \alpha_3 \cM_2 M_1 , \; L_4= M_4 - \alpha_4 \cM_2 M_2 + 
(\alpha_4-3) \cM_2^2 ,
\label{L3L4}
\eeq
where $\alpha_3$ and $\alpha_4$ are free parameters, one can easily see that
$\lag L_3\rag=\lag\Map^3\rag_c$ and $\lag L_4\rag=\lag\gammais^4\rag_c$
and that the variance of these estimators is minimum for :
\beq
\alpha_3= 3 + \frac{\lag\Map^4\rag_c}{\lag\Map^2\rag^2 (1+1/\rho)} ,
\label{alpha3}
\eeq
and:
\beqa
\lefteqn{ \!\! \alpha_4 = \frac{12\lag\gammais^2\rag^4 (1+1/\rho)^2 + \lag\gammais^4\rag_c \lag\gammais^2\rag^2 (14+8/\rho) + \lag\gammais^6\rag_c \lag\gammais^2\rag}{2\lag\gammais^2\rag^4 (1+1/\rho)^2 + \lag\gammais^4\rag_c \lag\gammais^2\rag^2}} \nonumber \\  
\label{alpha4}
\eeqa
Therefore, for a Gaussian distribution the estimators $L_3$ and $L_4$ show
the lowest scatter for $\alpha_3= 3$ and $\alpha_4=6$, that is when they are
identical to the estimators $H_3$ and $H_4$. Since weak lensing observables are
not exactly Gaussian, using $L_p$ instead of $H_p$ could further lower the 
dispersion of the measures. However, since the cumulants are not known a priori
(these are the quantities to be measured !) it is probably better to use the
simple estimators $H_p$. Moreover, as seen above the estimators $H_p$ have
the nice property to be always more efficient than $M_p$ whatever the actual
statistics of weak lensing observables. Another issue connected to the 
efficiency of weak lensing estimators is to select the optimal shape for the
filter $U_{\Map}(\vartheta)$ which defines the aperture mass. Such a study
is performed in Zhang et al. (2003). We shall not investigate this point 
further in this paper.

\section{Numerical results}
\label{Numerical results}

\subsection*{Cosmological model}

We now compute the statistics of weak lensing observables as they could be
measured from the SNAP mission. We focus on the low-order cumulants of the
aperture-mass and the shear as well as their pdf. We also consider the 
dispersion of the measures due to the intrinsic ellipticities of galaxies
and to the finite number of cells on the sky provided by the survey.

We consider a fiducial LCDM model with $\Om=0.3$, $\Ol=0.7$, $H_0=70$ 
km/s/Mpc and $\sigma_8=0.88$, in agreement with recent observations.
We shall also study in section~\ref{Wide} the effect of small variations
of these parameters onto weak-lensing observables.

The many-body correlations of the matter density field are obtained from
the minimal tree-model for the aperture-mass (Bernardeau \& Valageas 2000;
Munshi et al.2004) and the stellar model for the shear (Valageas et al.2004;
Munshi et al.2004), coupled to the fit to the non-linear power-spectrum
$P(k)$ of the dark matter density fluctuations given by Peacock \& Dodds 
(1996). These models are identical for the smoothed density field
and up to the third-order moment for any observable. From these density
correlations one obtains all cumulants of any weak-lensing observable as well
as its full pdf, as recalled in 
section~\ref{Redshift distribution of sources}. The predictions obtained
in this manner have been compared in details with results from numerical 
simulations in previous works (Bernardeau \& Valageas 2000;
Valageas et al.2004; Barber et al.2004; Munshi et al.2004) and have been seen
to provide good results.

\subsection*{Survey properties}

\begin{table}
\begin{center}
\caption{Survey parameters used for the numerical results, from the SNAP
mission as given in Refregier et al. (2004).}
\label{table1}
\begin{tabular}{lcccccc}
\hline
Survey & $A$ (deg$^2$) & $n_g$ (arcmin$^{-2}$) & $\sigma_*$ & $z_0$ \\
\hline
Deep & 15 & 260 & 0.36 & 1.31 \\
Wide & 300 & 100 & 0.31 & 1.13 \\
Wide+ & 600 & 68 & 0.30 & 1.07 \\
Wide- & 150 & 150 & 0.33 & 1.20 \\
\hline
\end{tabular}
\end{center}
\end{table}

Hereafter, we use the characteristics of the SNAP mission as given in 
Refregier et al.(2004), for several surveys. We recall their values in 
Table~1. The redshift distribution of galaxies is:
\beq
n(z_s) \propto z_s^2 \; e^{-(z_s/z_0)^2} \;\; \mbox{and} \;\; z_{\rm max}=3.
\label{nzSNAP}
\eeq
The shear variance due to intrinsic ellipticities and measurement errors is 
$\sigma_*=\lag|\epsilon_*|^2\rag^{1/2}$. The survey covers an area $A$ and 
the surface density of usable galaxies is $n_g$. Therefore, we take for the 
number $N$ of galaxies within a circular field of radius $\theta_s$:
\beq
N= n_g\pi\theta_s^2 \simeq 314 \left( \frac{n_g}{100 \mbox{arcmin}^{-2}} 
\right)  \left(\frac{\theta_s}{1\mbox{arcmin}}\right)^2,
\label{Nsurvey}
\eeq
and for the number $N_c$ of cells of radius $\theta_s$:
\beq
N_c= \frac{A}{(2\theta_s)^2} = 2.7 \times 10^5 \left(\frac{A}{300\mbox{deg}^2}
\right) \left(\frac{\theta_s}{1\mbox{arcmin}}\right)^{-2} .
\label{Ncsurvey}
\eeq
For the shear this number somewhat overestimates $N_c$ because of the 
sensitivity of $\gammais$ to long wavelengths, which would require 
the centres of different cells to be separated by more than $2\theta_s$ in 
order to be uncorrelated.

The SNAP mission will provide two surveys: a wide survey designed for weak
lensing with an area $A=300$ deg$^2$ and a deep survey with $A=15$ deg$^2$
designed for the search for Type Ia supernovae. We shall refer to them as
the ``Wide'' and ``Deep'' surveys. Following Refregier et al. (2004), 
in order to compare different survey strategies we shall
also study in section~\ref{Survey strategy} two hypothetical surveys, 
labeled ``Wide+'' and ``Wide-'' in Table~1, with the same observing time 
as the ``Wide'' survey (5 months) and a survey area $A$ which is doubled or 
halved (implying a smaller or larger depth at fixed observing time).
Finally, we also consider the two subsamples which can be obtained from the
``Wide'' survey by dividing galaxies into two redshift bins: $z_s>z_*$
(which we refer to as ``Wide$>$'') and $z<z_*$ (``Wide$<$''). We choose
$z_*=1.23$, which corresponds roughly to the separation provided by the SNAP
filters and which splits the ``Wide'' SNAP survey into two samples with the 
same number of galaxies (hence $n_g=50$ arcmin$^{-2}$).

\begin{figure}
\protect\centerline{
\epsfysize = 2.75truein
\epsfbox[18 140 688 715]
{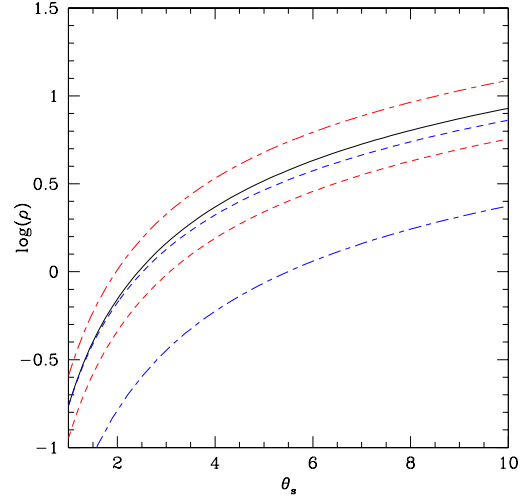}}
\caption{The ratio $\rho$ of cosmic variance over the noise due to the 
intrinsic ellipticity dispersion, from eq.(\ref{Mapc}), for the aperture-mass 
$\Map$. We show the results obtained for different surveys: ``Wide'' 
(solid line), ``Wide+'' (lower dashed line), ``Wide-'' (upper dot-dashed 
line), ``Wide$>$'' (upper dashed line) and ``Wide$<$'' (lower dot-dashed 
line).}
\label{figrho}
\end{figure}

We plot in Fig.~\ref{figrho} the ratio $\rho$ which measures the 
contribution to the noise from the ``cosmic variance'' relative to the
effect associated with the intrinsic ellipticity dispersion, from 
eq.(\ref{Mapc}), for the aperture-mass $\Map$. Of course, we can check that 
$\rho$ increases with the radius $\theta_s$ of the filter (i.e. with the 
number $N$ of galaxies within the circular field of radius $\theta_s$).
Thus, beyond a few arc-minutes the noise is dominated by the ``cosmic 
variance''. For the same reason, as compared with the fiducial ``Wide''
survey (solid line), $\rho$ is larger for the ``Wide-'' survey (upper 
dot-dashed line), which is narrower but deeper, and smaller for the ``Wide+''
survey (lower dashed line). On the other hand, we can see that $\rho$
is much smaller for the low-$z$ subsample ``Wide$<$'' (lower dot-dashed 
line), which contains fewer galaxies and has a lower variance $\lag\Map^2\rag$,
while it is almost unchanged for the high-$z$ subsample ``Wide$>$''
(upper dashed line) which has also twice fewer galaxies but a larger
variance $\lag\Map^2\rag$. As seen in section~\ref{Redshift binning}, it will
imply that although the skewness of the aperture-mass is smaller for the
high-$z$ subsample (and more difficult to measure) the deviations of the
pdf $\cP(M)$ from the Gaussian are easier to measure. The ratio $\rho$
obtained for the shear components shows similar behaviours.

\subsection{A specific case study: Wide SNAP survey}
\label{Wide}

We first consider the case of the Wide SNAP survey, the properties of which
are given in Table~1. We shall investigate the sensitivity to the survey
parameters in next sections.

\subsubsection{Variance}
\label{Variance}

\begin{figure}
\protect\centerline{
\epsfysize = 2.75truein
\epsfbox[18 140 688 715]
{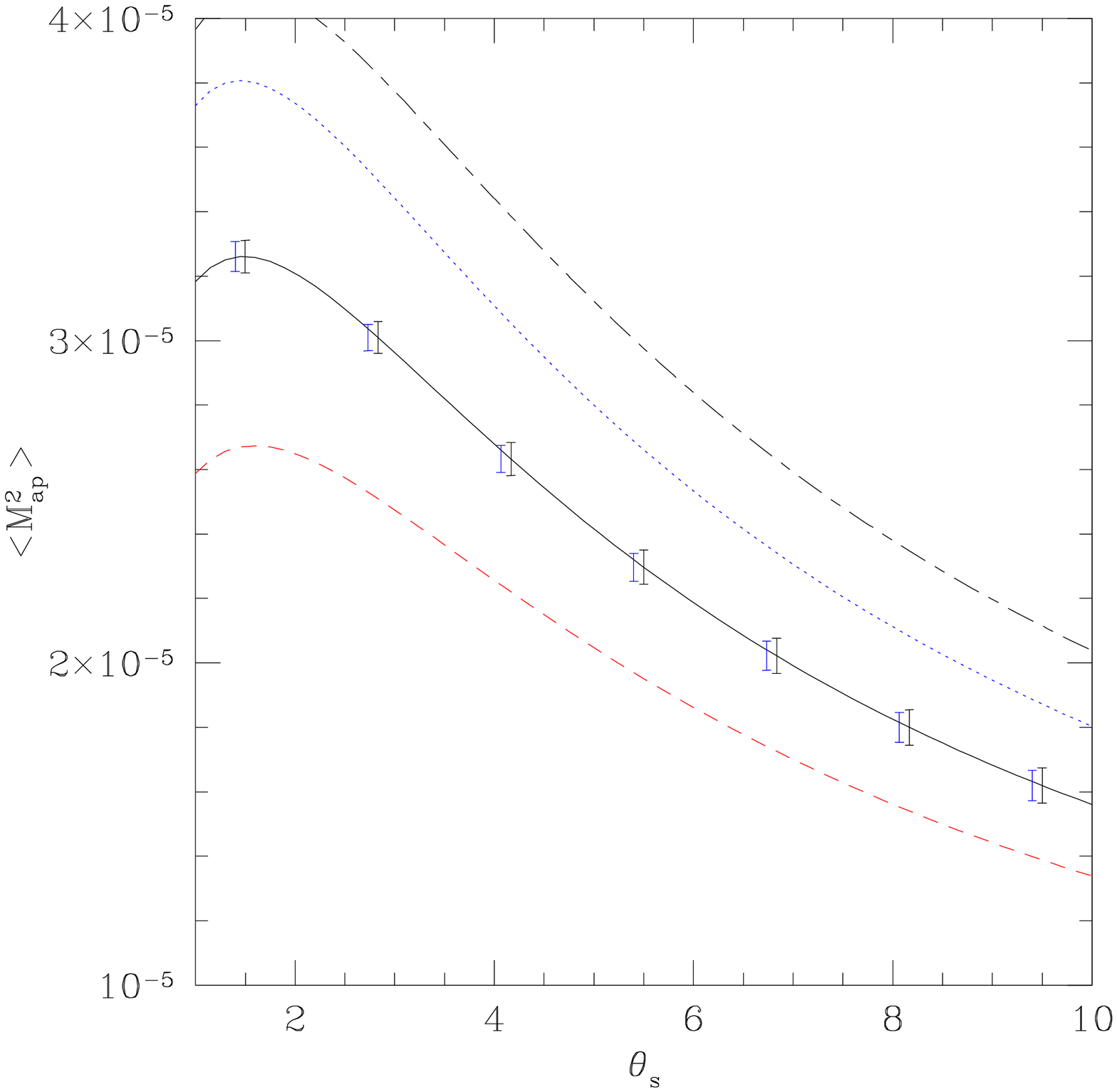}}
\caption{The variance $\lag\Map^2\rag$ of the aperture-mass (solid curve). 
The largest error bars show the $1-\sigma$ dispersion $\sigma(\cM_2)$ from 
eq.(\ref{sigM2}) and eq.(\ref{sigcMp}). The smaller error bars which are
slightly shifted to the left show the dispersion obtained by neglecting 
non-Gaussian contributions (i.e. $\lag\Map^4\rag_c=0$ in eq.(\ref{sigM2})). 
We also show the effect of a $10\%$ increase of $\Om$ (from $\Om=0.3$ up to 
$\Om=0.33$, central dotted curve), of a $10\%$ increase of the normalization 
$\sigma_8$ of the density power-spectrum (from $\sigma_8=0.88$ up to 
$\sigma_8=0.97$, upper dot-dashed curve), and of a $10\%$ decrease of the 
characteristic redshift $z_0$ (eq.(\ref{nzSNAP})) of the survey (from 
$z_0=1.13$ down to $z_0=1.02$, lower dashed curve).}
\label{figXiMap}
\end{figure}

\begin{figure}
\protect\centerline{
\epsfysize = 2.75truein
\epsfbox[18 140 688 715]
{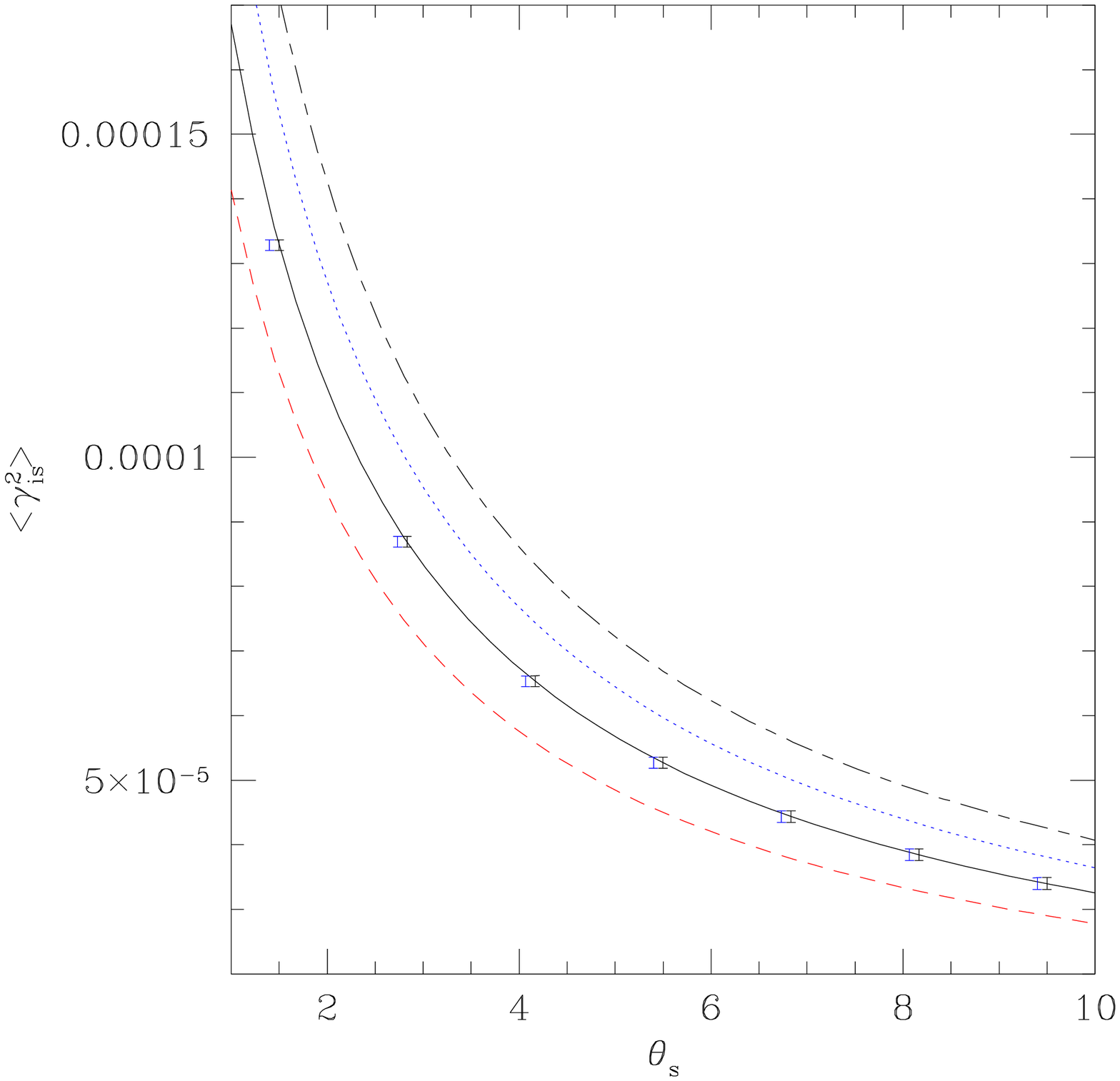}}
\caption{The variance $\lag\gammais^2\rag$ of the smoothed shear 
component $\gammais$ (solid curve). The largest error bars show the 
$1-\sigma$ dispersion $\sigma(\cM_2)$ from eq.(\ref{sigM2gamma}) and 
eq.(\ref{sigcMpgamma}). The smaller error bars which are slightly shifted to 
the left show the dispersion obtained by neglecting non-Gaussian 
contributions (i.e. $\lag\gammais^4\rag_c=0$ in eq.(\ref{sigM2gamma})). We 
also show the effect of a $10\%$ increase of $\Om$ (from $\Om=0.3$ up to 
$\Om=0.33$, central dotted curve), of a $10\%$ increase of the normalization 
$\sigma_8$ of the density power-spectrum (from $\sigma_8=0.88$ up to 
$\sigma_8=0.97$, upper dot-dashed curve), and of a $10\%$ decrease of the 
characteristic redshift $z_0$ (eq.(\ref{nzSNAP})) of the survey (from 
$z_0=1.13$ down to $z_0=1.02$, lower dashed curve).}
\label{figXigam}
\end{figure}

We show in Figs.~\ref{figXiMap}-\ref{figXigam} the variance of the 
aperture-mass $\lag\Map^2\rag$ and of the smoothed shear component
$\lag\gammais^2\rag$ for the wide SNAP survey. As recalled above, this
second-order moment is smaller for the aperture-mass $\Map$ which removes
the contribution of long-wavelength density fluctuations to weak-lensing.  
Moreover, it bends down for small angular scales $\theta_s \la 1'$ since in
this regime the projected density $\kappa({\vec \vartheta})$ shows more power
at larger scales (i.e. the non-linear density power-spectrum $P(k)$ grows more
slowly than $k^{-2}$ at high $k$). We also display the $1-\sigma$ dispersion 
$\sigma(\cM_2)$ from eq.(\ref{sigM2}) and eq.(\ref{sigcMp}) (largest error
bars in the figures). The smaller error bars which are slightly shifted to 
the left show the dispersion obtained by neglecting non-Gaussian 
contributions (i.e. $\lag\Map^4\rag_c=0$ in eq.(\ref{sigM2})). Thus, we can 
see that neglecting non-Gaussianities slightly underestimates the dispersion 
of the measurements but the difference with the full calculation is rather 
small. The relative size of the error bars is somewhat larger for the 
aperture-mass than for the shear because of the larger value of $\rho$. 
However, in both cases we can see that the wide 
survey of the SNAP mission should measure the variance of these weak-lensing 
observables up to a very good accuracy. We must point out, though, that this 
study does not take into account the possible systematic effects which might 
dominate the inaccuracy of the measures.

We also display the results obtained with a $10\%$ increase of $\Om$ 
(from $\Om=0.3$ up to $\Om=0.33$, central dotted curve), or a $10\%$ 
increase of the normalization $\sigma_8$ of the density power-spectrum 
(from $\sigma_8=0.88$ up to $\sigma_8=0.97$, upper dot-dashed curve), or a
$10\%$ decrease of the characteristic redshift $z_0$ (eq.(\ref{nzSNAP})) of 
the survey (from $z_0=1.13$ down to $z_0=1.02$, lower dashed curve). As is 
well-known and can be checked in Figs.~\ref{figXiMap}-\ref{figXigam}, the 
amplitude of gravitational lensing distortions increases with $\Om$ and
the matter density (see eq.(\ref{kappazs})), with the amplitude $\sigma_8$ of 
the density fluctuations (see eq.(\ref{kappazs})) and with the redshift $z_0$
as the line-of-sight is more extended. As seen in the figures, 
there is a clear degeneracy between these parameters. On the other hand, 
assuming other parameters are known we see that $\Om$ can be determined down 
to a few percents or to the relative accuracy of the redshift distribution
and half the relative accuracy of $\sigma_8$.

Here we must note that the different points in 
Figs.~\ref{figXiMap}-\ref{figXigam} are not fully independent since different 
wavelengths of the underlying density field are somewhat correlated. In order
to combine various angular scales to obtain an overall error-bar on a few 
cosmological parameters it is convenient to adopt a Fisher matrix approach
(e.g., Hu \& Tegmark 1999). However, we shall not investigate this traditional
approach here, as it has already been studied in the literature.

\subsubsection{Non-Gaussianities}
\label{Non-Gaussianities}

\begin{figure}
\protect\centerline{
\epsfysize = 2.75truein
\epsfbox[18 140 688 715]
{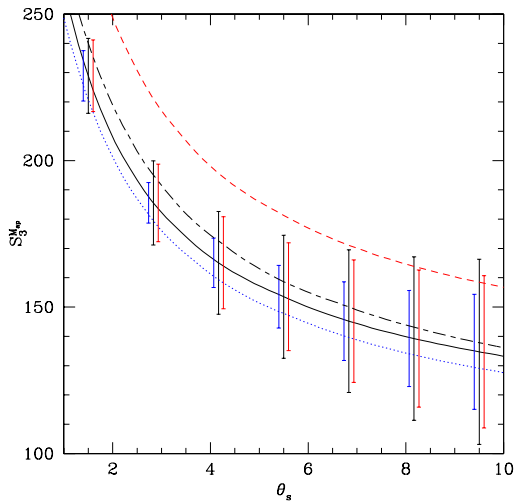}}
\caption{The skewness $S_3^{\Map}=\lag\Map^3\rag/\lag\Map^2\rag^2$ of the 
aperture-mass (solid curve). The central error bars show the $1-\sigma$ 
dispersion $\sigma(\cS_3)$ from eq.(\ref{cS3}). The smaller error bars which 
are slightly shifted to the left show the dispersion obtained by neglecting 
non-Gaussian contributions (i.e. $\lag\Map^6\rag_c=\lag\Map^4\rag_c=
\lag\Map^3\rag_c=0$ in eq.(\ref{sigM3})). The smaller error bars which are 
slightly shifted to the right show the dispersion obtained from the estimator 
$\cS_3^H$ in eq.(\ref{cS3H}). We also show the effect of a $10\%$ increase 
of $\Om$ (lower dotted curve), of a $10\%$ increase of $\sigma_8$ (central 
dot-dashed curve) and of a $10\%$ decrease of the redshift $z_0$ (upper 
dashed curve).}
\label{figSMap}
\end{figure}

\begin{figure}
\protect\centerline{
\epsfysize = 2.75truein
\epsfbox[18 140 688 715]
{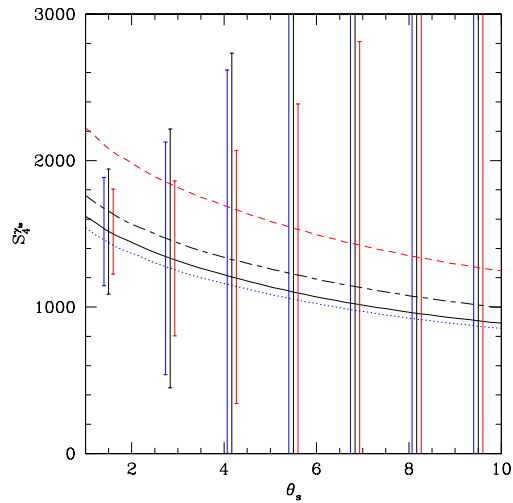}}
\caption{The kurtosis $S_4^{\gammais}=\lag\gammais^4\rag_c/
\lag\gammais^2\rag^3$ of the smoothed shear component. The central error 
bars show the $1-\sigma$ dispersion $\sigma(\cS_4)$ from eq.(\ref{cS4}). 
The smaller error bars which are slightly shifted to the left show the 
dispersion obtained by neglecting non-Gaussian contributions (i.e. 
$\lag\gammais^8\rag_c =\lag\gammais^6\rag_c=\lag\gammais^4\rag_c=0$ in 
eq.(\ref{sigM4gamma})). The smaller error bars which are slightly shifted 
to the right show the dispersion obtained from the estimator $\cS_4^H$ in 
eq.(\ref{cS4H}). We also show the effect of a $10\%$ increase of $\Om$
(lower dotted curve), of a $10\%$ increase of $\sigma_8$ (central dot-dashed
curve) and of a $10\%$ decrease of the redshift $z_0$ (upper dashed curve).}
\label{figSgam}
\end{figure}

Next, we display in Figs.~\ref{figSMap}-\ref{figSgam} the skewness 
$S_3^{\Map}$ of the aperture-mass and the kurtosis $S_4^{\gammais}$ of 
the shear component. These quantities provide a measure of the departure from
Gaussianity of the underlying matter density field. They can also be used
to break the degeneracy between the normalization of the density 
power-spectrum and the cosmological parameters (Bernardeau et al.1997).
We can check that the
error bars increase very fast with the order of the statistics. In particular,
it is clear that the dispersion is dominated by the error bars associated
with higher-order moments and we can neglect the dispersion due to the
denominators $\lag\Map^2\rag^2$ and $\lag\gammais^2\rag^3$ which enter 
the definition of the skewness and kurtosis.
The central error bars in the figures show the $1-\sigma$ dispersion
$\sigma(\cS_p)$ from eqs.(\ref{cS3}),(\ref{cS4}). It is much larger for the
shear kurtosis, which is a higher-order statistics, than for the aperture-mass
skewness. In particular, while the detection of non-Gaussianities from the
aperture-mass should be clear up to $10'$ and even somewhat beyond, it should
become rather difficult from the shear component for angular scales 
$\theta_s \ga 6'$. The smaller error bars which are slightly shifted to the 
left show the dispersion obtained by neglecting non-Gaussian contributions. 
Thus, neglecting
non-Gaussianities again leads to an underestimate of the dispersion of the
measures, but the effect remains small. On the other hand, the smaller error 
bars which are slightly shifted to the right show the dispersion obtained 
from the estimators $\cS_p^H$ in eqs.(\ref{cS3H}),(\ref{cS4H}). As seen in
section~\ref{Improving low-order estimators}, these estimators which directly
measure the cumulants always give a smaller dispersion than the estimators
$M_p$ which measure the moments. We can see in 
Figs.~\ref{figSMap}-\ref{figSgam} that the improvement is rather small for
the skewness of the aperture-mass but it is already significant for the
kurtosis of the shear. Therefore, these estimators should prove useful to 
extract quantitative informations from future weak-lensing surveys.

As for the variance, we also display the results obtained with a $10\%$ 
increase of $\Om$ (lower dotted curve), or a $10\%$ increase of $\sigma_8$ 
(central dot-dashed curve), or a $10\%$ decrease of the redshift $z_0$ 
(upper dashed curve). The skewness and the kurtosis show a modest dependence 
on the cosmological parameter (roughly of the same order: $8\%$) and they 
decrease for larger $\Om$. This can be understood from the factor $\Om$ 
which appears in eq.(\ref{kappazs}). They show a somewhat stronger dependence
on $\sigma_8$ and they actually increase with $\sigma_8$ (so that $\Om$
and $\sigma_8$ have opposite effects on the skewness and the kurtosis while
they acted in the same direction for the variance). This reflects the fact
that a higher $\sigma_8$ implies a matter density field which is deeper in
the non-linear regime and further away from the Gaussian. On the other hand, 
the skewness and the kurtosis show a strong variation with the redshift 
($\sim 17\%$ and $\sim 40\%$) and they increase for a smaller source redshift.
This can be understood from the fact that a larger source redshift means
a longer line-of-sight (whence the pdf becomes closer to a Gaussian as we
add the lensing contributions from successive mass sheets, following the
central limit theorem) and a matter density field which is closer to Gaussian.
Note that because the sensitivity onto $z_0$ and $\Om$ is different between
the second and higher-order moments, they can be used to constrain both
quantities and to remove the degeneracy between $z_0$ and $\Om$ which appeared
in the variance. On the other hand, assuming that the redshift distribution is
well known, higher order moments can also be used to break the degeneracy 
between the normalization $\sigma_8$ of the power-spectrum and the 
cosmological parameter $\Om$, as seen from the figures (also Bernardeau et 
al.1997). However, our results show that the error bar on the measure of 
$\Om$ cannot be smaller than twice the error bar on the redshift distribution.

\subsubsection{Probability distribution functions}
\label{Probability distribution functions}

\paragraph{Aperture-mass $\Map$}

\begin{figure}
\protect\centerline{
\epsfysize = 2.75truein
\epsfbox[18 120 688 715]
{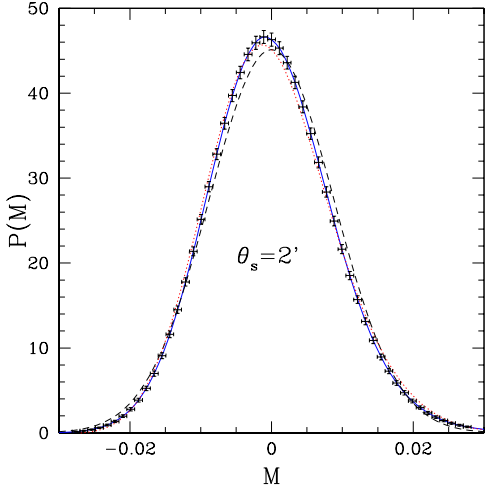}}
\caption{The pdf $\cP(M)$ from eq.(\ref{phiM}) for the estimator $M$ 
associated with the aperture-mass $\Map$ as defined in eq.(\ref{MapQ1}).
Note that the Gaussian noise introduced by intrinsic ellipticities makes
$\cP(M)$ closer to the Gaussian than the actual pdf $\cP(\Map)$ which
only takes into account gravitational lensing. The solid line shows the
theoretical prediction (\ref{phiM}), the dashed line is the Gaussian and
the dotted line is the Edgeworth expansion (\ref{Edg}) up to the first
non-Gaussian term (the skewness). The error bars show the $1-\sigma$
dispersion $\sigma(P_k)$ from eq.(\ref{sigPk}).}
\label{figPMapt2}
\end{figure}

\begin{figure}
\protect\centerline{
\epsfysize = 2.75truein
\epsfbox[18 120 688 715]
{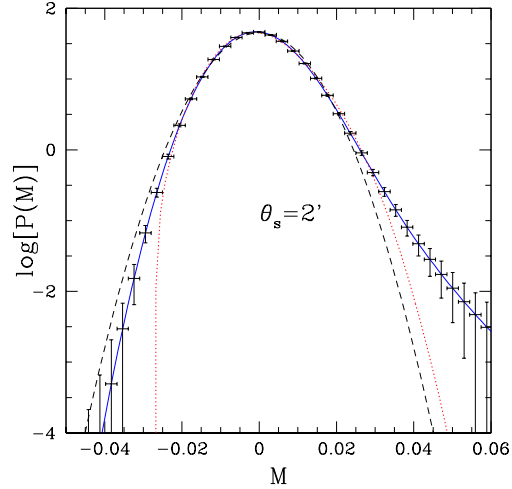}}
\caption{The logarithm $\log(\cP(M))$ of the pdf shown in 
Fig.~\ref{figPMapt2}. The error bars correspond to $\log(\cP(M)\pm\sigma)$,
from eq.(\ref{sigPk}).}
\label{figlPMapt2}
\end{figure}

Finally, we show in Figs.~\ref{figPMapt2}-\ref{figlPMapt2} the pdf $\cP(M)$
obtained for the estimator $M$ defined in eq.(\ref{MapQ1}). As seen in
section~\ref{Intrinsic ellipticity}, these pdfs are actually the convolution
of the pdf $\cP(\Map)$ (which measures gravitational lensing) by the Gaussian
of variance $\sigma_*^2$ associated with the noise due to the galaxy 
intrinsic ellipticity dispersion and to the detector white noise. This 
convolution makes the final pdf $\cP(M)$ closer to Gaussian than the actual
$\cP(\Map)$. We consider the angular scale $\theta_s=2'$ and we display the
theoretical prediction (\ref{phiM}) (solid line), the Gaussian (dashed line)
and the Edgeworth expansion (\ref{Edg}) up to the first non-Gaussian term 
(i.e. the skewness) (dotted line). We also show the $1-\sigma$ error bars
obtained from eq.(\ref{sigPk}). We chose for the width $\Delta$ of the 
intervals the values $\Delta=\lag M^2\rag^{1/2}/8$ in Fig.~\ref{figPMapt2}
and $\Delta=\lag M^2\rag^{1/2}/3$ in Fig.~\ref{figlPMapt2}. We can see from 
Fig.~\ref{figPMapt2} that it should be possible to measure the departure from
the Gaussian near the peak of $\cP(M)$, which is slightly higher and shifted
towards negative $M$. This also translates the asymmetry of $\cP(\Map)$.
On the other hand, from Fig.~\ref{figlPMapt2} it appears that the far 
positive tail of $\cP(M)$, for $M\simeq 0.05$, should also provide a means
to detect such non-Gaussianities. One should also be able to extract some
useful information from the negative tail at $M\simeq -0.03$. Note that it 
should be possible to distinguish $\cP(M)$ from both the Gaussian and the
Edgeworth expansion. This means that one has access to more information
than is encoded in the variance and the skewness. Thus, it would be 
interesting to check in future surveys that these three domains of $\cP(M)$ 
show the expected behaviour which characterizes the non-Gaussianities
brought by non-linear gravitational clustering. From another point of view, 
the expected shape of $\cP(M)$ due to gravitational lensing might be useful 
in order to discriminate this signal from possible non-Gaussianities induced 
by the detector (which might contaminate the measure of the skewness).

\paragraph{Smoothed shear component $\gammais$}

\begin{figure}
\protect\centerline{
\epsfysize = 2.75truein
\epsfbox[18 120 688 715]
{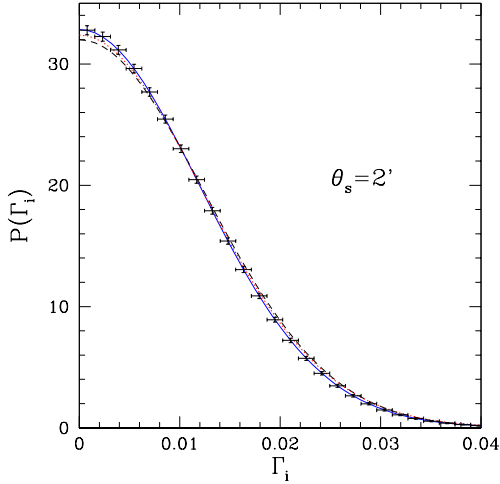}}
\caption{The pdf $\cP(\Gammai)$ from eq.(\ref{phiM}) for the estimator 
$\Gammai$ associated with the smoothed shear component $\gammais$ 
as defined in eq.(\ref{gammaQ1}). Note that the Gaussian noise introduced by 
intrinsic ellipticities makes $\cP(\Gammai)$ closer to the Gaussian 
than the actual pdf $\cP(\gammais)$ which only takes into account 
gravitational lensing. The solid line shows the theoretical prediction 
(\ref{phiM}), the dashed line is the Gaussian and the dotted line is the 
Edgeworth expansion (\ref{Edg}) up to the first non-Gaussian term (the 
kurtosis). The error bars show the $1-\sigma$ dispersion $\sigma(P_k)$ from 
eq.(\ref{sigPkgamma}).}
\label{figPgamt2}
\end{figure}

\begin{figure}
\protect\centerline{
\epsfysize = 2.75truein
\epsfbox[18 120 688 715]
{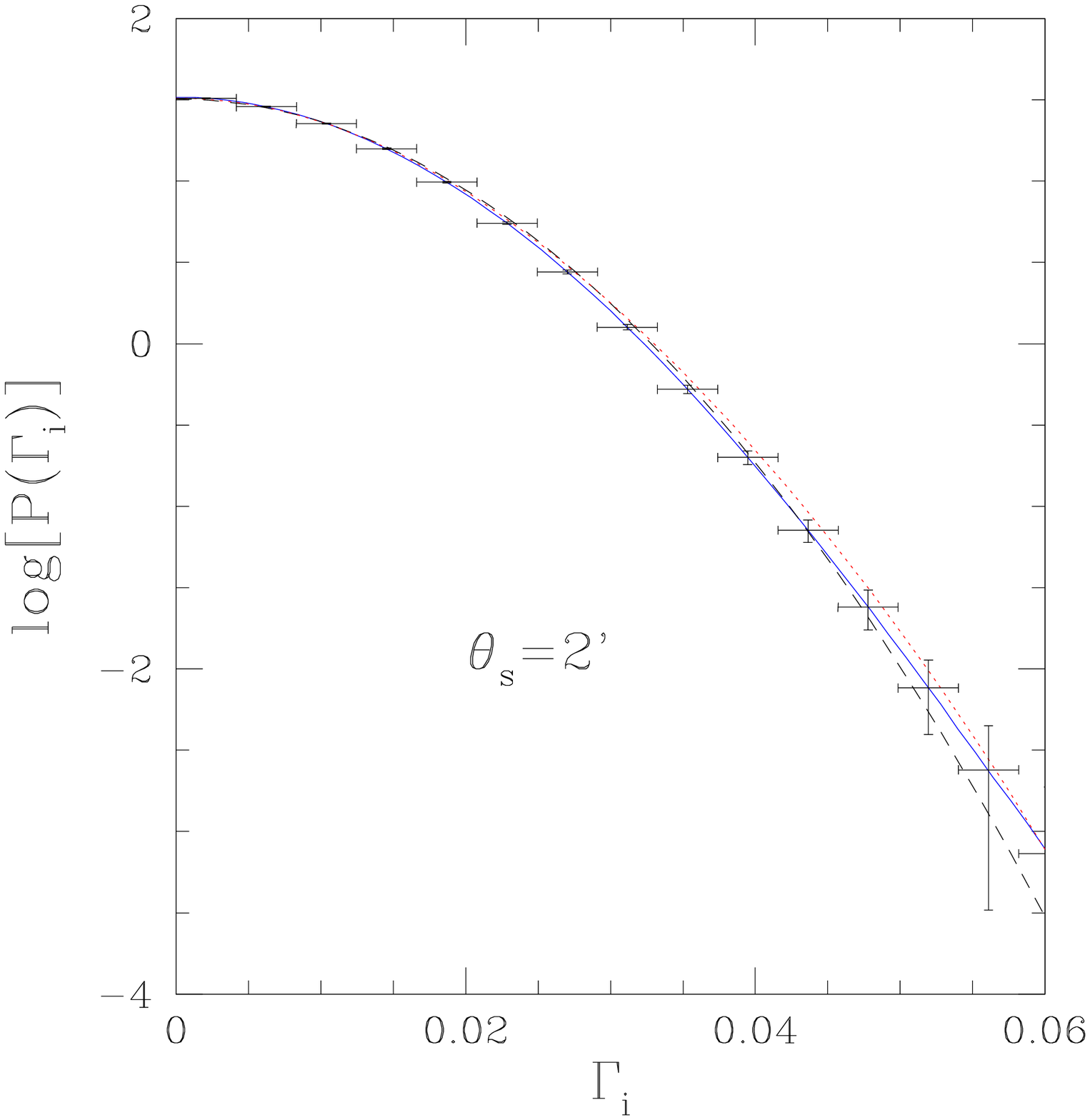}}
\caption{The logarithm $\log(\cP(\Gammai))$ of the pdf shown in 
Fig.~\ref{figPgamt2}. The error bars correspond to 
$\log(\cP(\Gammai)\pm\sigma)$, from eq.(\ref{sigPkgamma}).}
\label{figlPgamt2}
\end{figure}

We show in Figs.~\ref{figPgamt2}-\ref{figlPgamt2} the pdf $\cP(\Gammai)$
obtained for the estimator $\Gammai$ defined in eq.(\ref{gammaQ1}).
As for the aperture-mass, this is actually the convolution of 
$\cP(\gammais)$ by a Gaussian of variance $\sigma_*^2$, which models the
noise due to galaxy intrinsic ellipticities and detector white noise. We
again display the theoretical prediction (\ref{phiM}) (solid line), the
Gaussian (dashed line) and the Edgeworth expansion (\ref{Edg}) up to the 
first non-Gaussian term (i.e. the kurtosis). The $1-\sigma$ error bars are
obtained from eq.(\ref{sigPkgamma}) with $\Delta=\lag\Gammai^2\rag^{1/2}/8$ 
in Fig.~\ref{figPgamt2} and $\Delta=\lag\Gammai^2\rag^{1/2}/3$ in 
Fig.~\ref{figlPgamt2}. We see from Fig.~\ref{figPgamt2}
that it should again be possible to measure the deviations from Gaussianity
near the peak of $\cP(\Gammai)$ ($\Gammai\simeq 0$), which is 
slightly higher than for a Gaussian. The non-Gaussianity might also be measured
from the near tail of $\cP(\Gammai)$, at $\Gammai\simeq 0.02$.
As for the aperture-mass, measuring the pdf $\cP(\Gammai)$ in two
different ranges provides useful information since it can be used to check
the shape of the non-Gaussianities expected from non-linear gravitational
clustering, or to discriminate the signal from spurious non-Gaussianities
introduced by the detector. Note indeed that one should be able to
distinguish $\cP(\Gammai)$ from both the Gaussian and the Edgeworth 
expansion. On the other hand, we see from Fig.~\ref{figlPgamt2} that the far 
tail of the pdf ($\Gammai\ga 0.04$) is too noisy to give useful 
constraint on non-Gaussianities, contrary to the aperture-mass.

\subsection{Survey strategy: Width vs. Depth}
\label{Survey strategy}

We have seen in the previous sections that the nominal Wide SNAP survey 
should yield useful information about the amplitude and the non-Gaussianities
of the matter density fluctuations. We now study the dependence of these
results on the survey properties. Thus, following Refregier et al. (2004),
we compare this Wide SNAP survey with the Deep survey realized by the same 
mission (designed for the search for Type Ia supernovae) and with two 
hypothetical surveys, labeled ``Wide+'' and ``Wide-'' in Table~1, with the
same observing time (5 months) and a survey area $A$ which is doubled or 
halved (implying a smaller or larger depth at fixed observing time).

\subsubsection{Variance}
\label{Variance all}

\begin{figure}
\protect\centerline{
\epsfysize = 2.75truein
\epsfbox[18 140 688 715]
{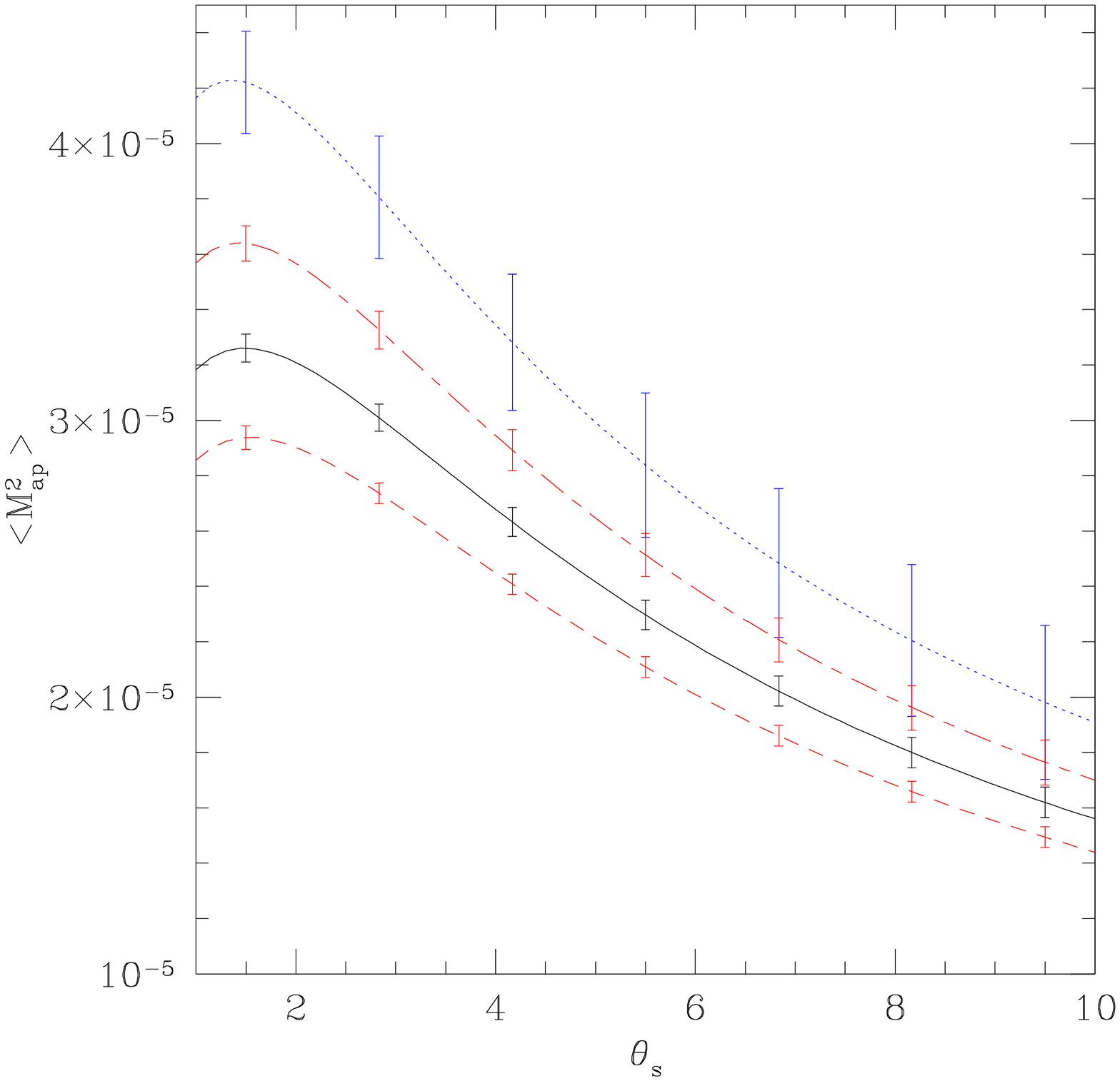}}
\caption{The variance $\lag\Map^2\rag$ of the aperture-mass for the Wide
survey (solid line), the Deep survey (dotted line), the ``Wide+'' survey
(lower dashed line) and the ``Wide-'' survey (upper dot-dashed line). 
The error bars show the $1-\sigma$ dispersion $\sigma(\cM_2)$ from 
eq.(\ref{sigM2}) and eq.(\ref{sigcMp}).}
\label{figXiMapall}
\end{figure}

\begin{figure}
\protect\centerline{
\epsfysize = 2.75truein
\epsfbox[18 140 688 715]
{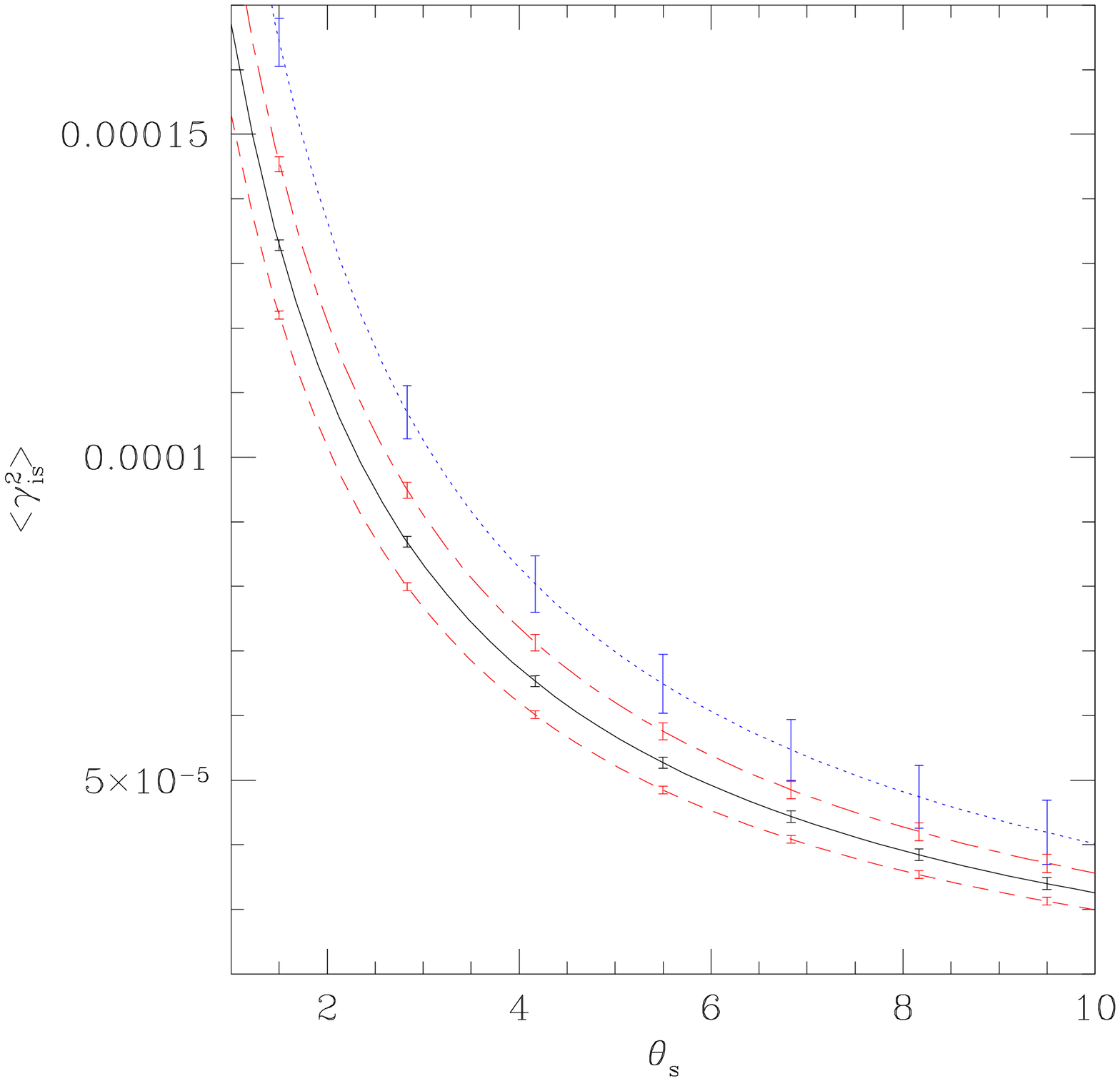}}
\caption{The variance $\lag\gammais^2\rag$ of the smoothed shear 
component $\gammais$ for the Wide survey (solid line), the Deep survey 
(dotted line), the ``Wide+'' survey (lower dashed line) and the ``Wide-'' 
survey (upper dot-dashed line). The error bars show the $1-\sigma$ 
dispersion $\sigma(\cM_2)$ from eq.(\ref{sigM2}) and eq.(\ref{sigcMp}).}
\label{figXigamall}
\end{figure}

We show in Figs.~\ref{figXiMapall}-\ref{figXigamall} the variance of the 
aperture-mass $\lag\Map^2\rag$ and of the smoothed shear component
$\lag\gammais^2\rag$ for these four surveys: ``Wide'' (solid line, as in
section~\ref{Variance}), ``Deep'' (dotted line), ``Wide+'' (dashed line) and
``Wide-'' (dot-dashed line). Of course, the variance is larger for the Deep
survey since the redshift distribution is broader. However, its error bars are
larger because the total survey area is much smaller. In agreement with 
Refregier et al. (2004), we find that the widest survey ``Wide+'' yields
slightly smaller error bars than the nominal survey ``Wide'' (or the
narrower survey ``Wide-''). Hence a wider and shallower survey is slightly
more efficient but the difference is probably too small to have any impact
on observational strategies.

\subsubsection{Non-Gaussianities}
\label{Non-Gaussianities all}

\begin{figure}
\protect\centerline{
\epsfysize = 2.75truein
\epsfbox[18 140 688 715]
{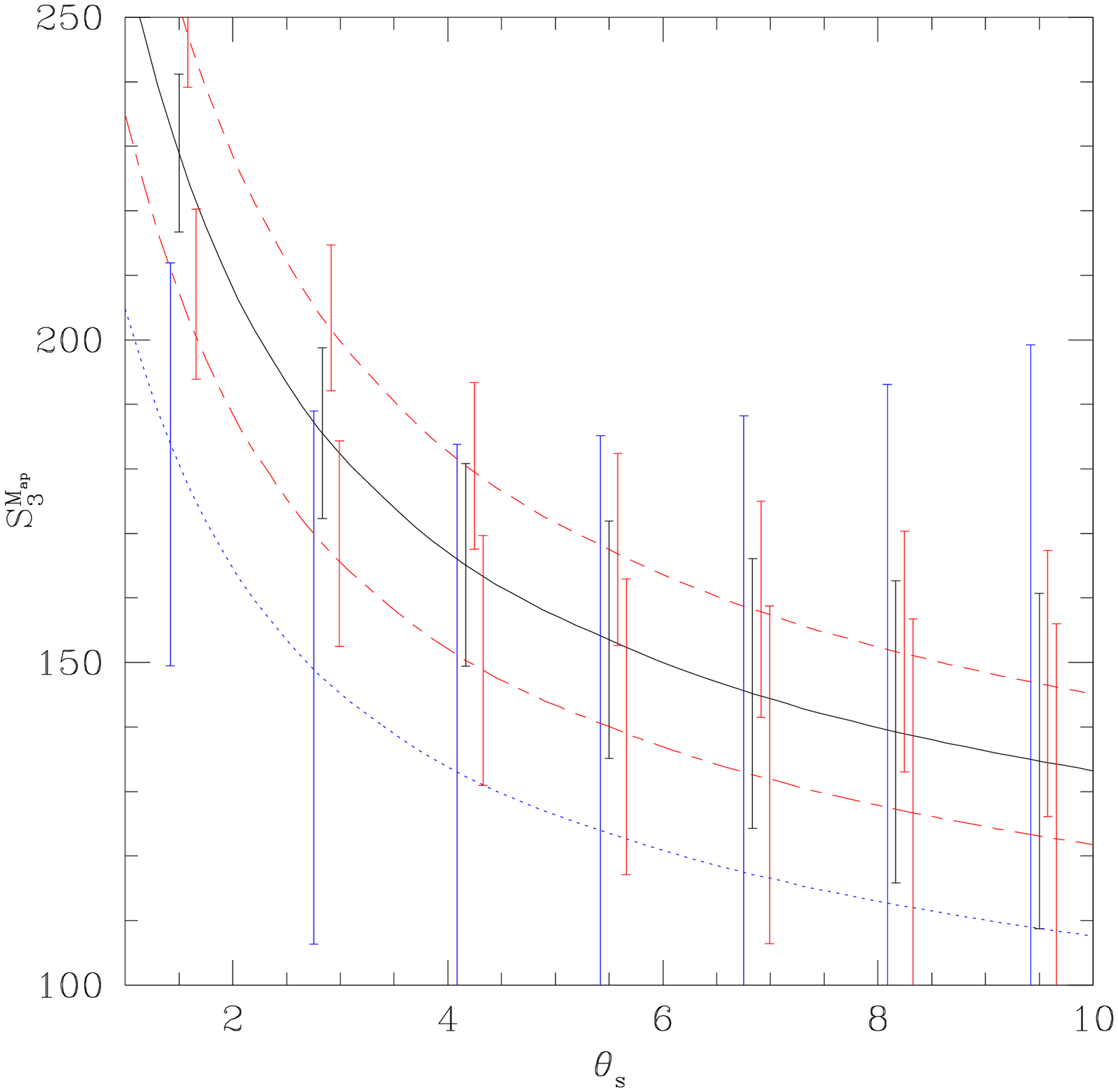}}
\caption{The skewness $S_3^{\Map}$ of the aperture-mass for the Wide survey 
(solid line), the Deep survey (dotted line), the ``Wide+'' survey (upper 
dashed line) and the ``Wide-'' survey (lower dot-dashed line). The error bars 
show the dispersion obtained from the estimator $\cS_3^H$ in eq.(\ref{cS3H}).}
\label{figSMapall}
\end{figure}

\begin{figure}
\protect\centerline{
\epsfysize = 2.75truein
\epsfbox[18 140 688 715]
{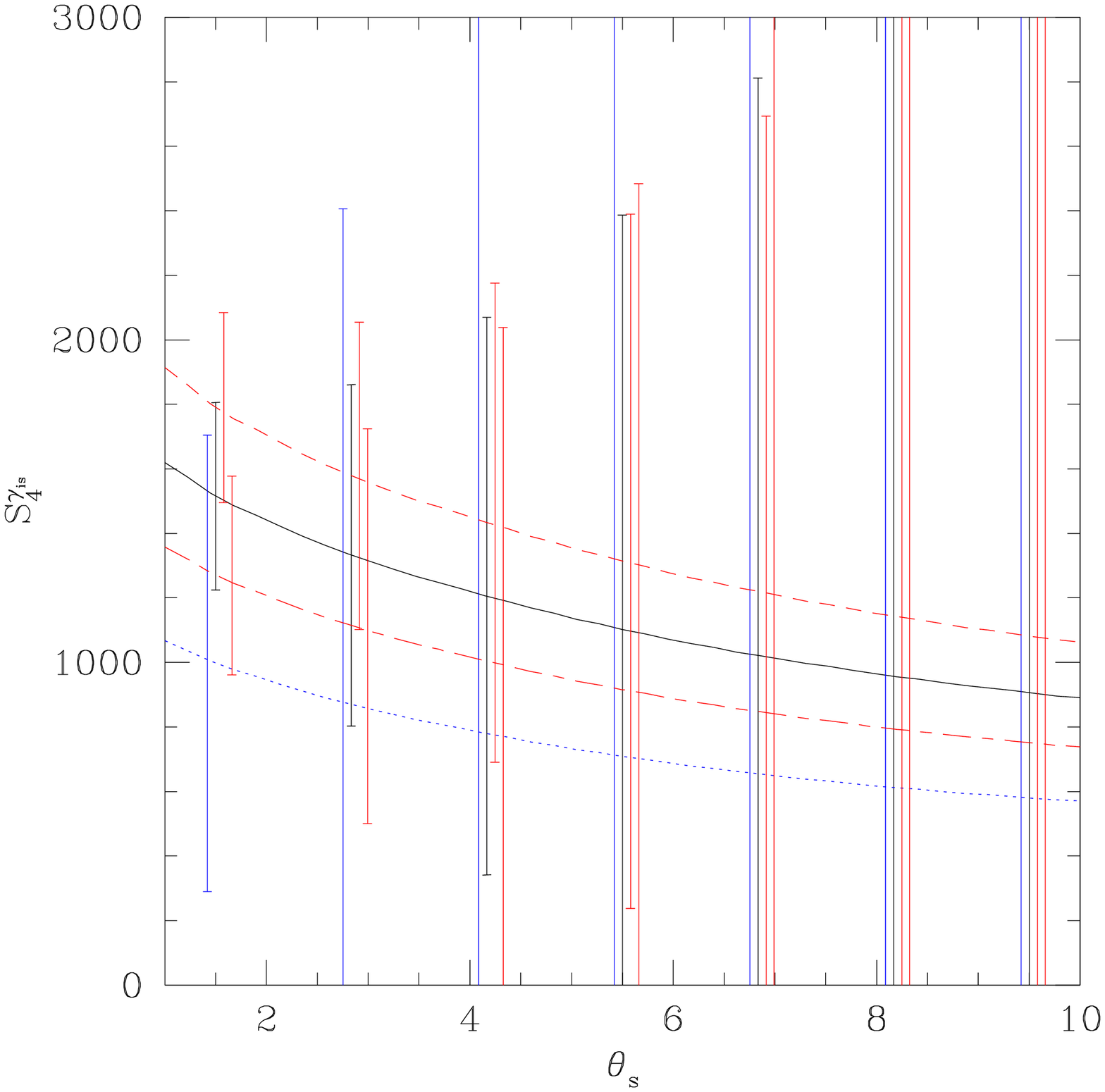}}
\caption{The kurtosis $S_4^{\gammais}$ of the smoothed shear component for 
the Wide survey (solid line), the Deep survey (dotted line), the ``Wide+'' 
survey (upper dashed line) and the ``Wide-'' survey (lower dot-dashed line).
The error bars show the dispersion obtained from the estimator $\cS_4^H$ in 
eq.(\ref{cS4H}).}
\label{figSgamall}
\end{figure}

Next, we display in Figs.~\ref{figSMapall}-\ref{figSgamall} the skewness 
$S_3^{\Map}$ of the aperture-mass and the kurtosis $S_4^{\gammais}$ of 
the shear component for the four surveys. The error bars correspond to the
estimators $H_p$ which show less scatter than $M_p$. The skewness and the 
kurtosis are smaller for the ``Deep'' and ``Wide-'' surveys which have a 
redshift distribution of sources which is more heavily weighted by high 
redshifts. As for the variance, the ``Wide+'' survey yields the best results, 
since it exhibits larger non-Gaussianities and smaller error bars. Thus, it 
enables one to detect non-Gaussianities up to slightly larger angular scales 
than the nominal ``Wide'' survey would allow.

\subsection{Survey strategy: Redshift binning}
\label{Redshift binning}

In the previous sections, we have described how the quality of the 
information obtained from weak lensing measures depend on the survey 
properties. However, once a specific survey is realized one still has the
possibility to analyze the data in different ways, for instance by subdividing
the galaxy sample into several redshifts bins. This can be conveniently
done by using photometric redshifts. To investigate this issue, we describe
in this section the results which can be obtained by dividing the Wide SNAP
survey, given in Table~1, into two redshifts bins: $z_s>z_*$ (which we denote
by ``Wide$>$'') and $z<z_*$ (which we refer to as ``Wide$<$''). We choose 
$z_*=1.23$, which corresponds roughly to the separation provided by the SNAP
filters and which splits the Wide SNAP survey into two samples with the same
number of galaxies (hence $n_g=50$ arcmin$^{-2}$).

\subsubsection{Variance}
\label{Variance z}

\begin{figure}
\protect\centerline{
\epsfysize = 2.75truein
\epsfbox[18 140 688 715]
{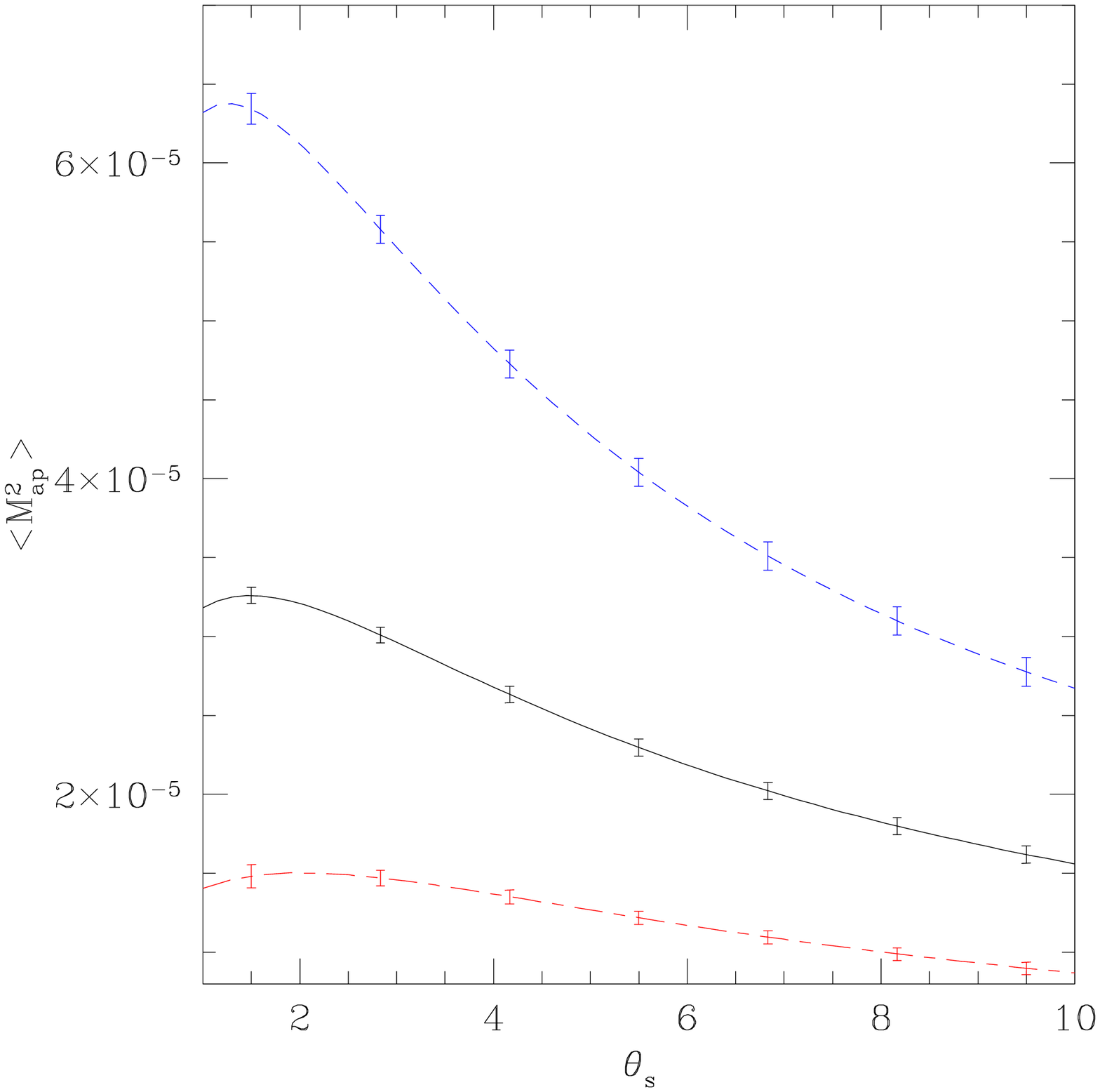}}
\caption{The variance $\lag\Map^2\rag$ of the aperture-mass for the full Wide
survey (solid line), the high-$z$ ``Wide$>$'' sample (upper dashed line) and 
the low-$z$ ``Wide$<$'' sample (lower dot-dashed line). The error bars show 
the $1-\sigma$ dispersion $\sigma(\cM_2)$ from eq.(\ref{sigM2}) and 
eq.(\ref{sigcMp}).}
\label{figXiMapz}
\end{figure}

\begin{figure}
\protect\centerline{
\epsfysize = 2.75truein
\epsfbox[18 140 688 715]
{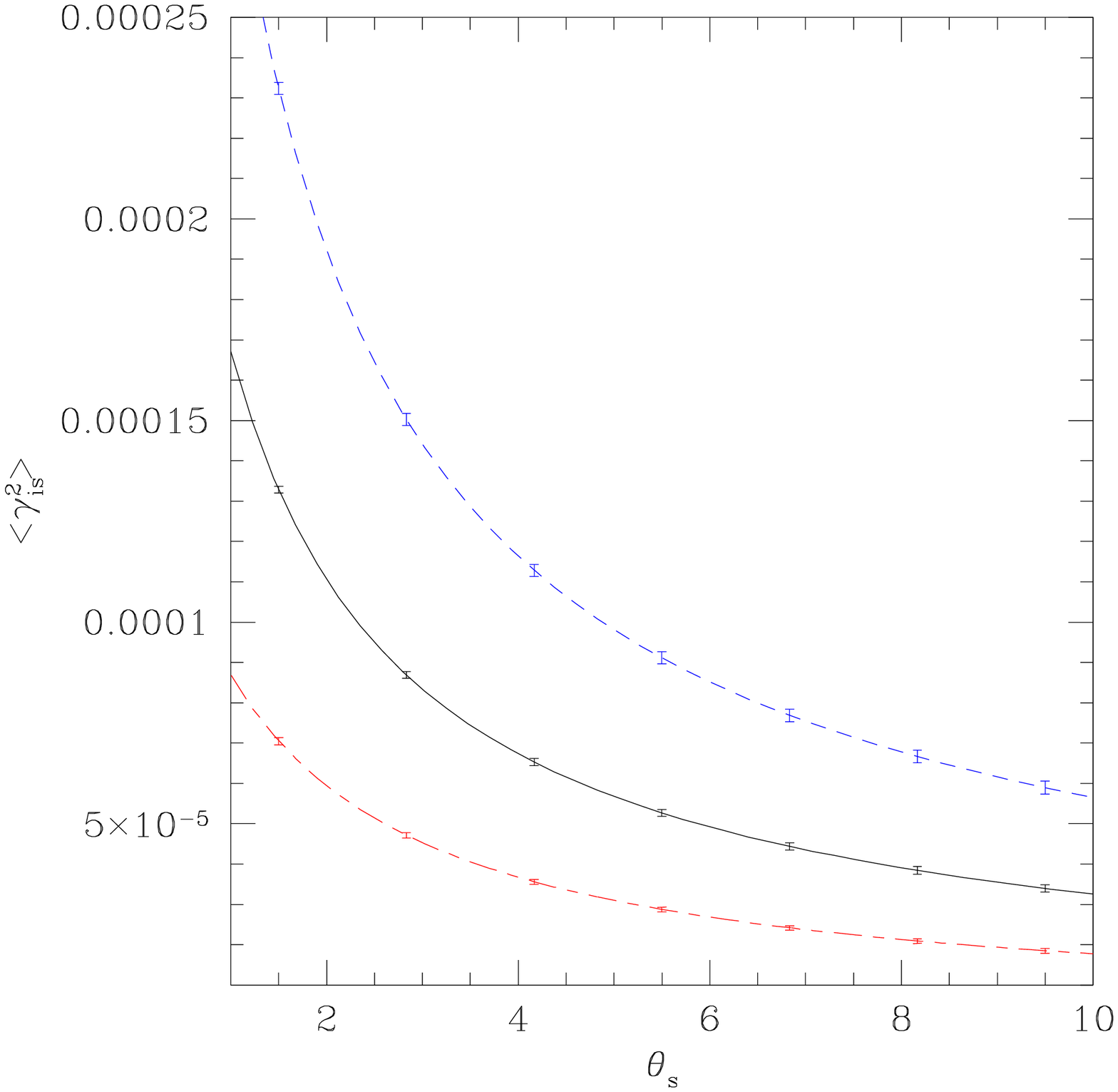}}
\caption{The variance $\lag\gammais^2\rag$ of the smoothed shear 
component $\gammais$ for the full Wide survey (solid line), the high-$z$ 
``Wide$>$'' sample (upper dashed line) and the low-$z$ ``Wide$<$'' sample 
(lower dot-dashed line). The error bars show the $1-\sigma$ dispersion 
$\sigma(\cM_2)$ from eq.(\ref{sigM2}) and eq.(\ref{sigcMp}).}
\label{figXigamz}
\end{figure}

We show in Figs.~\ref{figXiMapz}-\ref{figXigamz} the variance of the 
aperture-mass and of the smoothed shear component for the three samples: the
full Wide SNAP survey (solid line), the high-$z$ ``Wide$>$'' sample (upper 
dashed line) and the low-$z$ ``Wide$<$'' sample (lower dot-dashed line).
Of course, we find that the variance is larger for the high-$z$ sample
since the amplitude of weak lensing distortions increases with the redshift
of the source (and the length of the line-of-sight). Note that the error bars
are quite small for all three samples therefore it is interesting to divide
the survey into several redshift bins which allow one to check the evolution
with time of the matter power-spectrum. This also provides stronger 
constraints on cosmological parameters (and possibly on the equation of state
of the dark energy, which we shall not investigate here).

We must note that the different curves in 
Figs.~\ref{figXiMapz}-\ref{figXigamz} show some correlation since the lines 
of sight to distant sources located in different redshift bins probe the same
density fluctuations at low $z$. Again it can be convenient to use a Fisher 
matrix approach to combine these various redshifts. We shall study the 
cross-correlations between different redshift subsamples in a future paper
(Munshi \& Valageas 2004).

\subsubsection{Non-Gaussianities}
\label{Non-Gaussianities z}

\begin{figure}
\protect\centerline{
\epsfysize = 2.75truein
\epsfbox[18 140 688 715]
{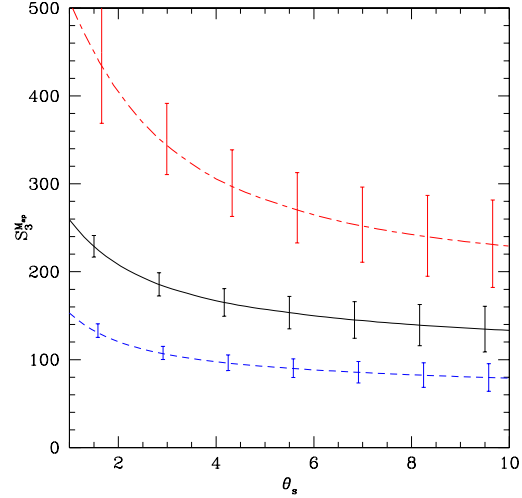}}
\caption{The skewness $S_3^{\Map}$ of the aperture-mass for the full Wide 
SNAP survey (solid line), the high-$z$ ``Wide$>$'' sample (lower 
dashed line) and the low-$z$ ``Wide$<$'' sample (upper dot-dashed line). The 
error bars show the dispersion obtained from the estimator $\cS_3^H$ in 
eq.(\ref{cS3H}).}
\label{figSMapz}
\end{figure}

\begin{figure}
\protect\centerline{
\epsfysize = 2.75truein
\epsfbox[18 140 688 715]
{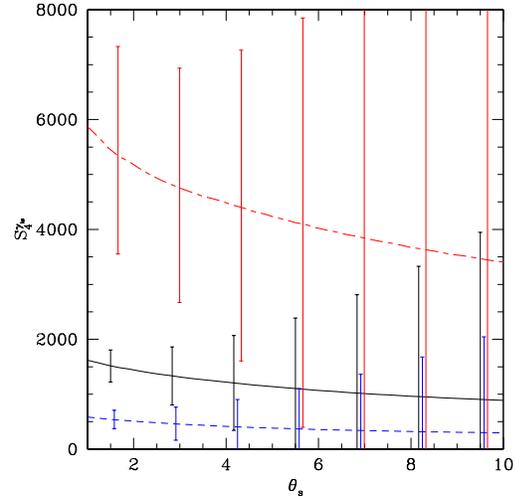}}
\caption{The kurtosis $S_4^{\gammais}$ of the smoothed shear component for 
the full Wide survey (solid line), the high-$z$ ``Wide$>$'' sample
(lower dashed line) and the low-$z$ ``Wide$<$'' sample (upper dot-dashed 
line). The error bars show the dispersion obtained from the estimator 
$\cS_4^H$ in eq.(\ref{cS4H}).}
\label{figSgamz}
\end{figure}

Next, we show in Figs.~\ref{figSMapz}-\ref{figSgamz} the skewness 
$S_3^{\Map}$ of the aperture-mass and the kurtosis $S_4^{\gammais}$ of the 
smoothed shear component for the three samples. The parameters $S_p$ are 
smaller for the high-$z$ sample (lower dashed line) which involves the 
convolution of the weak lensing effects arising from many successive mass 
sheets (which makes the signal closer to Gaussian, following the central limit 
theorem) and which probes a density field which is closer to the linear 
Gaussian regime. In the case of the aperture-mass, all three samples allow
a clear detection of non-Gaussianity and a rather precise measure of 
$S_3^{\Map}$. As for the variance, it will be interesting to perform such
a redshift binning of the data in order to check the evolution with redshift
of $S_3^{\Map}$. This should strengthen the constraints obtained from 
observations. Moreover, it could be useful in order to discriminate the 
non-Gaussianities due to the non-linear gravitational dynamics from those
associated with the noise which might be non-Gaussian (whether it comes from
the galaxy intrinsic ellipticities or the detector itself). In the case of the
shear component the three samples enable one to detect non-Gaussianity at small
angular scales $\theta_s \la 4'$ while the low-$z$ sample allows one to go
up to slightly larger angles, $\theta_s \la 7'$, because it yields a kurtosis
which is much larger. Note that in both cases, the skewness or the kurtosis
shows a strong variation with the redshift binning, which should easily be
detected.

\subsubsection{Probability distribution functions}
\label{Probability distribution functions z}

\begin{figure}
\protect\centerline{
\epsfysize = 2.75truein
\epsfbox[18 120 688 715]
{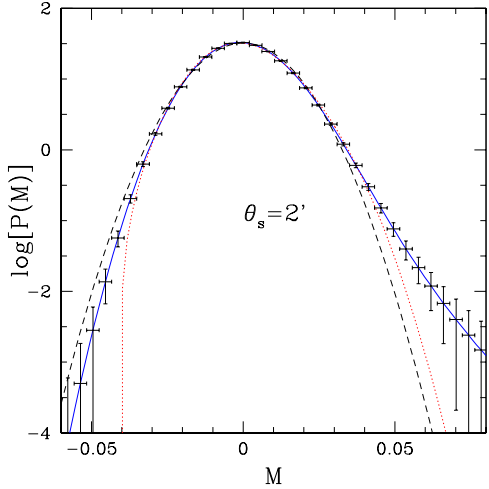}}
\caption{The logarithm $\log(\cP(M))$ of the pdf $\cP(M)$ from 
eq.(\ref{phiM}) for the high-$z$ sample ``Wide$>$'', for the estimator $M$ 
associated with the aperture-mass $\Map$ as defined in eq.(\ref{MapQ1}). 
The solid line shows the theoretical prediction (\ref{phiM}), the dashed 
line is the Gaussian and the dotted line is the Edgeworth expansion 
(\ref{Edg}) up to the first non-Gaussian term (the skewness). The error 
bars correspond to $\log(\cP(M)\pm\sigma)$, from eq.(\ref{sigPk}).}
\label{figlPMapt2zp}
\end{figure}

Finally, we display in Fig.~\ref{figlPMapt2zp} the pdf $\cP(M)$ obtained
for the estimator $M$ associated with the aperture-mass $\Map$ for the
high-$z$ sample ``Wide$>$''. Indeed, although the skewness $S_3^{\Map}$ is
larger for the low-$z$ sample, its pdf $\cP(M)$ is closer to Gaussian because
the variance $\lag\Map^2\rag$ is smaller so that the influence of the 
intrinsic galaxy ellipticities is larger and this turns out to be the main
factor. Therefore, we find that for the low-$z$ sample the pdf $\cP(M)$ cannot
be easily distinguished from the Gaussian. By contrast, as seen in 
Fig.~\ref{figlPMapt2zp} the high-$z$ sample still allows a clear detection of
non-Gaussianity. In fact, as for the full sample studied in 
section~\ref{Probability distribution functions}, the tails of the distribution
enable one to distinguish $\cP(M)$ from both the Gaussian and the Edgeworth
expansion. This should again prove useful. On the other hand, the pdf
$\cP(\Gammai)$ cannot be distinguished from the Gaussian, except near its
peak for the full sample.

\section{Discussion}


Weak lensing surveys are already 
being used to constrain allowed regions of cosmological
parameter space. Future surveys such as SNAP will provide
a better opportunity by covering a large fraction of the sky.
While there has been a tremendous progress in understanding
the effect of cosmological parameters on weak lensing statistics,
a complete analysis of realistic noise contribution 
for various survey strategies is still lacking.

In this paper we have mainly focused on realistic surveys
such as SNAP to compute the estimator induced variance
due to contributions from the finite catalogue size and the intrinsic 
ellipticity distribution
of galaxies. Although Poisson effects (due to the discrete distribution of
galaxies) are quite small for such surveys because of the high surface 
density of galaxies $n_g$, the other contributions can play a dominant role.
We study both the aperture-mass $\Map$ and the smoothed shear components 
$\gammais$, which can be more easily recovered from actual surveys than
the smoothed convergence $\kappa_{\rm s}$. Extending earlier works which
focused on the lower order cumulants, we present a unified approach in 
order to handle both cumulants and the full pdfs of these objects.

In agreement with previous works (Refregier et al. 2004), we find that
surveys like the SNAP mission can measure the variance of both $\Map$ and 
$\gammais$ up to a very good accuracy (a few percent) for the entire range 
of angular scales $1'<\theta_s<10'$ that we have studied. This should yield
strong constraints on the cosmological parameters (e.g., a measure of $\Om$
to a few percent if all other parameters are known). However, there is a
well-known degeneracy between $\Om$ and $\sigma_8$. In addition, we find
that $\Om$ cannot be measured to better than the relative accuracy of the
redshift distribution, which might be a significant limitation.

As usual, the degeneracy between several cosmological parameters can be
removed by measuring higher-order moments. We find that the skewness 
$S_3^{\Map}$ of the aperture-mass should be easily detected and measured up
to a $10\%$ accuracy over $1'<\theta_s<10'$. By contrast, the shear kurtosis
$S_4^{\gammais}$ should be difficult to measure beyond $6'$. 
Indeed, higher-order cumulants
are increasingly difficult to measure from noisy data: their scatter grows
with their order as a larger number of terms contributes to their dispersion
which also involves all cumulants up to twice their order. Using a realistic
model for the underlying matter density field, our computation of these
error bars takes into account all these cumulants (that is we do not keep
only the Gaussian terms or multiply this contribution by a fudge factor),
which slightly increase the dispersion. On the other hand, we introduced
a new class of estimators $\cH_p$ designed to measure these low-order
cumulants. We have shown that they yield a scatter which is always smaller
than the one obtained by using the simple estimators $\cM_p$ derived from the 
moments themselves (and are actually optimal among a one-parameter family 
of estimators for a Gaussian distribution). We find that the gain is rather 
small for the skewness $S_3^{\Map}$ but for the kurtosis $S_4^{\gammais}$ 
we get a significant reduction of the scatter. Since these estimators are 
no more difficult to use than the naive $\cM_p$ estimators they should be
preferred over the latter. The skewness or the kurtosis may be used to remove
the degeneracy between $\Om$ and $\sigma_8$ so as to enable one to measure
the cosmological parameters. However, we note that they are very sensitive to
the redshift of the sources, so that the accuracy of $\Om$ cannot be smaller
than twice the error bar on the redshift of the galaxy distribution. On the
other hand, by binning the sample over redshift one might be able to 
discriminate the influence of the redshift.
 
In addition to the low-order moments we have also studied the full pdfs
$\cP(M)$ and $\cP(\Gammai)$, where $M$ (resp. $\Gammai$) is a biased
estimator of the aperture-mass (resp. of the shear component). Here the
intrinsic galaxy ellipticity distribution plays a key role as it makes
$M$ and $\Gammai$ biased estimators and it makes the pdfs $\cP(M)$ and 
$\cP(\Gammai)$ much closer to the Gaussian than the pdfs $\cP(\Map)$ and
$\cP(\gammais)$. Note that the intrinsic galaxy ellipticity also contributed
to the measure of the low-order cumulants but it was less of a problem because
one can still build unbiased estimators of these low-order cumulants (by 
counting each galaxy only once, which also removes some contributions to their
scatter). We find that the pdfs $\cP(M)$ and $\cP(\Gammai)$ can be 
distinguished from a Gaussian through their shape near their maximum. 
Moreover, the negative and positive tails of the pdf $\cP(M)$ associated with
the aperture-mass can be discriminated from both the Gaussian and the
Edgeworth expansion (using only the first non-Gaussian term defined by the
skewness). This means that one can extract more information than is encoded
in the first two low-order moments. Moreover, by measuring the pdf $\cP(M)$
over these three domains one should be able to strengthen the constraints
derived for the underlying matter density field and to discriminate possible
non-Gaussianities induced by the detector. On the other hand, we find that
the far tails of the symmetric pdf $\cP(\Gammai)$ associated with the shear 
component are too noisy to give useful constraints. A detailed $\chi^2$
analysis will be presented elsewhere for simulated observations.

Next, we have investigated whether the information obtained from observations
could be improved by changing the survey strategy. Thus, we have compared the
nominal wide SNAP survey with the ``Deep'' SNAP survey as well as with two
hypothetical surveys with the same observing time as a the original survey
but a different trade-off between area and depth. We focused on the first two
low-order moments for the aperture-mass and the shear component. In agreement
with Refregier et al. (2004) (who studied the lensing power-spectrum and the
convergence skewness) we find that a wider and shallower survey is slightly
more efficient.

Finally, we have studied the possibility to divide the wide SNAP survey into
two redshift bins ($z<1.23$ and $z>1.23$). All three samples provide a very
accurate measure of the variance of both the aperture-mass and the shear
component. As noticed above, this should allow one the improve the constraints
and to check the evolution with redshift of non-linear gravitational 
clustering. The skewness $S_3^{\Map}$ is also well measured in the three 
samples, while the kurtosis $S_4^{\gammais}$ is more easily obtained from the
low-$z$ sample. This shows again the interest of such a redshift binning of 
the data. In addition, the high-$z$ sample again allows a good measure of
the pdf $\cP(M)$ associated with the aperture-mass, which can be distinguished
from both the Gaussian and the Edgeworth expansion.


In this article we have mainly focused on errors associated with 
quantities derived from a specific redshift bin. We shall extend our studies 
to the cross correlations among various redshift bins in future works
(Munshi \& Valageas 2004). 
This can also be generalized to compute the cross correlations among various
statistics derived from different surveys with non-identical scan strategies.


Throughout our studies we have ignored source clustering and the effect due 
to lens coupling.
A detailed prediction of source clustering and lens coupling
will require an accurate picture of how galaxy number densities
are related to the underlying mass distribution. 
Some of these issues have been studied by Bernardeau (1997),
Bernardeau et al. (1997) and Schneider et al. (1998) who found
that such corrections are negligible at least in the quasi-linear
regime. In the non-linear regime one might use numerical simulations
in order to evaluate this affect, but this would again require a specific
recipe for the correlation between galaxies and dark matter.


Another ingredient in our calculations has been the so called 
Born approximation. Its validity can only be checked numerically in the highly
non-linear regime. Thus, the consistency of analytical results and
numerous numerical simulations found in previous studies strongly suggests
that this approximation remains accurate in the highly non-linear regime
(see also Vale \& White 2003).


In our studies we also assumed that the intrinsic ellipticities of different
galaxies are uncorrelated (Heymans \& Heavens 2003, Crittenden et al. 2001)
(but of course we take into account its variance). Techniques 
to deal with such correlations have been studied in detail although the 
extent to which such correlations will affect weak lensing surveys remains 
somewhat uncertain. It is however generally believed that such
effects will play a less important role as we increase the survey depth
and we can reduce their role through acquisition of photometric redshift 
(Heymans \& Heavens 2003).


Although almost all present studies assume these intrinsic ellipticities to
be Gaussian random variables this might not be the case. Any signature of 
such non-Gaussianity if found by observational teams will have to be folded 
into analytical calculations. This can be performed in a straightforward way
within our formalism. In this case, the measure of full pdfs like $\cP(M)$
or $\cP(\Gammai)$ would be of great interest in order to disantengle the
signal from the non-Gaussianities due to galaxy ellipticities (which might
contaminate low-oder moments, especially if there are cross-correlations). 
However, if such intrinsic non-Gaussianities are too large they might
dominate the signal and preclude an accurate measure of the non-Gaussianities
due to the matter density field. In addition to the galaxy intrinsic 
ellipticities and to the finite size of the survey, another source of noise
is given by the finite PSF effects (smear). It would be interesting to include
this contribution in future studies.

\section*{acknowledgments}

DM was supported by PPARC of grant
RG28936. It is a pleasure for DM to acknowledge many fruitful
discussions with members of Cambridge Leverhulme Quantitative
Cosmology Group. This work has been 
supported by PPARC and the numerical work carried out with 
facilities provided by the University of Sussex. AJB was
supported in part by the Leverhulme Trust.


\begin{thebibliography}{}
\bibitem{} Bacon D.J., Refregier A., Ellis R.S., 2000, MNRAS, 318, 625
\bibitem{} Barber A. J., Munshi D., Valageas P., 2004, MNRAS, 347, 665
\bibitem{} Bernardeau F., Kofman L., 1995, ApJ, 443, 479
\bibitem{} Bernardeau F., Schaeffer R., 1992, A\&A, 255, 1
\bibitem{} Bernardeau F., Valageas P., 2000, A\&A, 364, 1
\bibitem{} Bernardeau F., Van Waerbeke L., Mellier Y., 1997, A\&A, 322, 1
\bibitem{CBT} Couchman H. M. P., Barber A. J., Thomas P. A., 1999,
MNRAS, 308, 180
\bibitem{}  Crittenden R.G., Natarajan P., Pen U., Theuns T., 2001, 
ApJ, 559, 552
\bibitem{} Cooray A., Sheth R., 2002, Phys.Rept., 372 ,1
\bibitem{} Fry J.N., 1984, ApJ, 279, 499
\bibitem{} Heymans C., Heavens A., 2003, MNRAS, 339, 711
\bibitem{} Hoekstra H., Yee H. K. C., Gladders M. D., 2002, ApJ, 577, 595
\bibitem{} Hu W., Tegmark M., 1999, ApJ, 514, L65
\bibitem{} Jain B., Seljak U., 1997, ApJ, 484, 560
\bibitem{} Jain B., Seljak U., White S.D.M., 2000, ApJ, 530, 547
\bibitem{} Jaroszynski M., Park C., Paczynski B., Gott J.R.,
1990, ApJ, 365, 22
\bibitem{} Kaiser N., 1992, ApJ, 388, 272
\bibitem{} Kaiser N., Squires G., Fahlman G., Woods D., 1994, in: Durret F., 
Mazure A., Tran Thanh Van J. (eds.) Clusters of galaxies. 
Editions Frontieres
\bibitem{} Kaiser N., 1998, ApJ, 498, 26
\bibitem{} Munshi D.,  2000, MNRAS, 318, 145
\bibitem{} Munshi D., Jain B., 2000, MNRAS, 318, 109
\bibitem{} Munshi D., Jain B., 2001, MNRAS, 322, 107
\bibitem{} Munshi D., Coles P., 2003, MNRAS, 338, 846
\bibitem{} Munshi D., Valageas P., 2004, astro-ph/0406081
\bibitem{} Munshi D., Melott A.L., Coles P., 1999, MNRAS, 311, 149
\bibitem{} Munshi D., Valageas P., Barber A. J., 2004, MNRAS, 350, 77
\bibitem{} Peacock J.A., Dodds S.J., 1996, MNRAS, 280, L19
\bibitem{} Peacock J.A., Smith R. E., 2000, MNRAS, 318, 1144
\bibitem{} Refregier A. et al., 2004, AJ, 127, 3102
\bibitem{} Schaeffer R., 1984, A\&A, 134, L15
\bibitem{} Schneider P., 1996, MNRAS, 283, 837
\bibitem{} Schneider P., van Waerbeke L., Jain B., Kruse G., 1998, MNRAS, 
296, 873
\bibitem{} Schneider P., Weiss A., 1988, ApJ, 330,1
\bibitem{} Szapudi I., Szalay A.S., 1993, ApJ, 408, 43
\bibitem{} Szapudi I., Szalay A.S., 1997, ApJ, 481, L1
\bibitem{} Takada M., Jain B., 2002, MNRAS, 337, 875
\bibitem{} Takada M., Jain B., 2003a, MNRAS, 344, 857
\bibitem{} Takada M., Jain B., 2003b, MNRAS, 346, 949
\bibitem{} Valageas P., 2000a, A\&A, 354, 767
\bibitem{} Valageas P., 2000b, A\&A, 356, 771
\bibitem{} Valageas P., Barber A. J., Munshi D., 2004, MNRAS, 347, 652
\bibitem{} Vale C., White M., 2003, ApJ, 592, 699 
\bibitem{} Van Waerbeke L., Bernardeau F., Mellier Y., 1999,
A\& A, 342, 15
\bibitem{} Van Waerbeke L., Hamana T., Scoccimarro R., Colombi S.,
Bernardeau F., 2001, MNRAS, 322, 918
\bibitem{} Van Waerbeke L., Mellier Y., Pell\'o R., Pen U.-L., McCracken
H.J., Jain B., 2002, A\&A, 393, 369
\bibitem{} Villumsen J.V., 1996, MNRAS, 281, 369
\bibitem{} Wambsganss J., Cen R., Ostriker J.P., 1998, ApJ, 494, 29
\bibitem{} Zhang T.-J., Pen U.-L., Zhang P., Dubinski J., 2003, ApJ, 598, 818



\end{thebibliography}
\end{document}